\documentclass[twocolumn]{IEEEtran}
\IEEEoverridecommandlockouts
\usepackage{amsmath,amssymb,amscd,latexsym,dsfont}
\usepackage{float}
\usepackage{color}
\usepackage{graphicx}
\usepackage{comment}
\usepackage{subfigure}
\usepackage{enumerate}
\usepackage{stfloats}
\usepackage{psfrag}
\usepackage{setspace}
\usepackage{balance}
\usepackage{multirow}
\usepackage{algorithmic}
\usepackage[]{algorithm}
\usepackage{pst-node}
\usepackage{xcolor}
\usepackage{cite}

\newcommand{\LOEP}[1]{\textbf{#1}}

\renewcommand{\markboth}[1]
  {\renewcommand{\leftmark}{#1}\renewcommand{\rightmark}{#1}}
\title{On Optimal TCM Encoders}

\author{
	\IEEEauthorblockN{
	Alex Alvarado, \emph{Member, IEEE},
	Alexandre Graell i Amat, \emph{Senior Member, IEEE}, \\
	Fredrik Br\"annstr\"om, \emph{Member, IEEE}, and
	Erik Agrell, \emph{Senior Member, IEEE}
	\thanks{Parts of this work were presented at the Information Theory and Applications (ITA) Workshop, San Diego, CA, February 2012, and at the IEEE International Symposium on Information Theory (ISIT) 2012, Cambridge, MA, July 2012.}
	\thanks{Research supported by the European Community's Seventh's Framework Programme (FP7/2007-2013) under grant agreement No. 271986, by the Swedish Research Council under grants \#621-2006-4872 and \#621-2011-5950, and by the Swedish Agency for Innovation Systems (VINNOVA) under the P3660$4$-1 MAGIC project. The calculations were performed
	on resources provided by the Swedish National Infrastructure
	for Computing (SNIC) at C3SE.}
	\thanks{A. Alvarado is with the Department of
	Engineering, University of Cambridge, UK (email:
	alex.alvarado@ieee.org). A. Graell i Amat, F. Br\"annstr\"om and Erik Agrell are with the
	Department of Signals and Systems, Chalmers University of Technology, Gothenburg,
	Sweden (email: \{alexandre.graell,fredrik.brannstrom,
	agrell\}@chalmers.se).}
	}
}

\newcommand{\dE}{d}
\newcommand{\Ad}{A_{d^2}}
\newcommand{\tAd}{\tilde{A}_{\tilde{d}^2}}
\newcommand{\Bd}{B_{d^2}}
\newcommand{\tBd}{\tilde{B}_{\tilde{d}^2}}

\newcommand{\Awdl}{A_{w,d^2,\ell}}
\newcommand{\Adl}{A_{d^2,\ell}}

\newcommand{\T}{^{\mathsf{T}}}            		
\newcommand{\tr}[1]{\mathrm{#1}}

\newcommand{\mc}[1]{\mathcal{#1}}

\newcommand{\set}[1]{\{#1\}}
\newcommand{\cd}{\cdot}
\newcommand{\vd}{\vdots}
\newcommand{\ld}{\ldots}
\newcommand{\dd}{\ddots}
\newcommand{\ms}[1]{\mathds{#1}}

\newcommand{\un}[1]{\underline{#1}}

\newcommand{\ie}{i.e.,~}
\newcommand{\eg}{e.g.,~}

\newcommand{\cf}{cf.~}
\newtheorem{theorem}{Theorem}

\newtheorem{lemma}{Lemma}
\newtheorem{definition}{Definition}

\newtheorem{example}{Example}

\newcommand{\bn}{\boldsymbol{n}}

\newcommand{\br}{\boldsymbol{r}}
\newcommand{\bs}{\boldsymbol{x}}
\newcommand{\bg}{\boldsymbol{g}}

\newcommand{\bX}{\boldsymbol{X}}

\newcommand{\bY}{\boldsymbol{Y}}

\newcommand{\bc}{\boldsymbol{c}}
\newcommand{\mcX}{\mc{X}}
\newcommand{\mcB}{\mc{B}}

\newcommand{\mcR}{\mc{R}}
\newcommand{\mcT}{\mc{T}}

\newcommand{\bb}{\boldsymbol{b}}
\newcommand{\bx}{\boldsymbol{x}}
\newcommand{\bi}{\boldsymbol{i}}
\newcommand{\bj}{\boldsymbol{j}}

\newcommand{\bu}{\boldsymbol{u}}
\newcommand{\bv}{\boldsymbol{v}}
\newcommand{\bz}{\boldsymbol{z}}
\newcommand{\by}{\boldsymbol{y}}

\newcommand{\matX}{{\mathop{\boldsymbol{X}}}}
\newcommand{\matG}{{\mathop{\boldsymbol{G}}}}
\newcommand{\matN}{{\mathop{\boldsymbol{N}}}}
\newcommand{\matB}{{\mathop{\boldsymbol{B}}}}

\newcommand{\matL}{{\mathop{\boldsymbol{L}}}}
\newcommand{\matT}{{\mathop{\boldsymbol{T}}}}
\newcommand{\matS}{{\mathop{\boldsymbol{X}}}}

\newcommand{\Es}{E_\tr{s}}
\newcommand{\Ns}{N_\tr{s}}

\renewcommand{\exp}[1]{\mathrm{exp}{\left(#1\right)}}
\newcommand{\tabref}[1]{Table~\ref{#1}}
\newcommand{\figref}[1]{Fig.~\ref{#1}}
\newcommand{\secref}[1]{Sec.~\ref{#1}}
\newcommand{\ind}{\mathit{index}}
\newcommand{\pointer}{\mathit{pointer}}

\begin{document}
\maketitle

\begin{abstract}
An asymptotically optimal trellis-coded modulation (TCM) encoder requires the joint design of the encoder and the binary labeling of the constellation. Since analytical approaches are unknown, the only available solution is to perform an exhaustive search over the encoder and the labeling. For large constellation sizes and/or many encoder states, however, an exhaustive search is unfeasible. Traditional TCM designs overcome this problem by using a labeling that follows the set-partitioning principle and by performing an exhaustive search over the encoders. In this paper we study binary labelings for TCM and show how they can be grouped into classes, which considerably reduces the search space in a joint design. For $8$-ary constellations, the number of different binary labelings that must be tested is reduced from $8!=40320$ to $240$. For the particular case of an $8$-ary pulse amplitude modulation constellation, this number is further reduced to $120$ and for $8$-ary phase shift keying to only $30$. An algorithm to generate one labeling in each class is also introduced. Asymptotically optimal TCM encoders are tabulated which are up to $0.3$~dB better than the previously best known encoders.
\end{abstract}

\section{Introduction}\label{Sec:Introduction}

The first breakthrough in coding for the bandwidth-limited regime came with Ungerboeck's trellis-coded modulation (TCM) \cite{Ungerboeck76,Ungerboeck82,Ungerboeck87a,Ungerboeck87b} in the early 80s where the concept of labeling by set-partitioning (SP) was introduced. TCM was quickly adopted in the modem standards in the early 90s and is a well studied topic \cite{Biglieri91_Book}, \cite[Sec.~8.12]{Proakis08_Book}, \cite[Ch.~18]{Lin04_Book}. Another important discovery in coded modulation (CM) design came in 1992 when Zehavi introduced the so-called bit-interleaved coded modulation (BICM) \cite{Zehavi92,Caire98}, usually referred to as a pragmatic approach for CM \cite{Fabregas08_Book}.

The design philosophies behind TCM and BICM for the additive white Gaussian noise (AWGN) channel are quite different. Ungerboeck's scheme is constructed coupling together a convolutional encoder and a constellation labeled using the SP principle. For constellations having certain symmetries, SP can be achieved by using the natural binary code (NBC) \cite[Fig.~4]{Ungerboeck82}, \cite[Fig.~3]{Ungerboeck87b}. On the other hand, BICM is typically a concatenation of a convolutional encoder and a constellation labeled by the binary reflected Gray code (BRGC) \cite{Gray53,Agrell04} through a bit-level interleaver. The BRGC is often used in BICM because it maximizes the BICM generalized mutual information for medium and high signal-to-noise ratios \cite[Sec.~III]{Caire98}, \cite[Sec.~IV]{Alvarado11b}. In TCM, the selection of the convolutional encoder is done so that the minimum Euclidean distance (MED) is maximized, while in BICM the encoders are the ones optimized for binary transmission.  BICM systems are then based on maximum free Hamming distance codes \cite[Sec.~12.3]{Lin04_Book} or on the so-called optimum distance spectrum (ODS) encoders first tabulated in \cite[Tables~III--V]{Chang97} and \cite[Tables~II--IV]{Bocharova97} and later extended in \cite{Frenger99}.

\begin{figure}
\newcommand{\scale}{0.8}
\psfrag{a}[cc][cc][\scale]{(a)}
\psfrag{b}[cc][cc][\scale]{(b)}
\psfrag{c}[cc][cc][\scale]{(c)}
\psfrag{bi}[Br][Br][\scale]{$i_{1,n}$}
\psfrag{x}[Bl][Bl][\scale]{$x_n$}
\psfrag{T}[Bl][Bl][\scale]{Transform}
\psfrag{SP}[Bl][Bl][\scale]{SP Mapper (NBC)}
\psfrag{BRGC}[Bl][Bl][\scale]{BRGC Mapper}
\psfrag{CC1}[Bl][Bl][\scale]{$\matG=[13,4]$}
\psfrag{CC2}[Bl][Bl][\scale]{$\matG=[13,17]$}
\psfrag{x1}[cc][cc][\scale]{$x_1$}
\psfrag{x2}[cc][cc][\scale]{$x_2$}
\psfrag{x3}[cc][cc][\scale]{$x_3$}
\psfrag{x4}[cc][cc][\scale]{$x_4$}
\psfrag{l1}[cc][cc][\scale]{$00$}
\psfrag{l2}[cc][cc][\scale]{$01$}
\psfrag{l3}[cc][cc][\scale]{$10$}
\psfrag{l4}[cc][cc][\scale]{$11$}
\begin{center}
	\includegraphics[width=0.95\columnwidth]{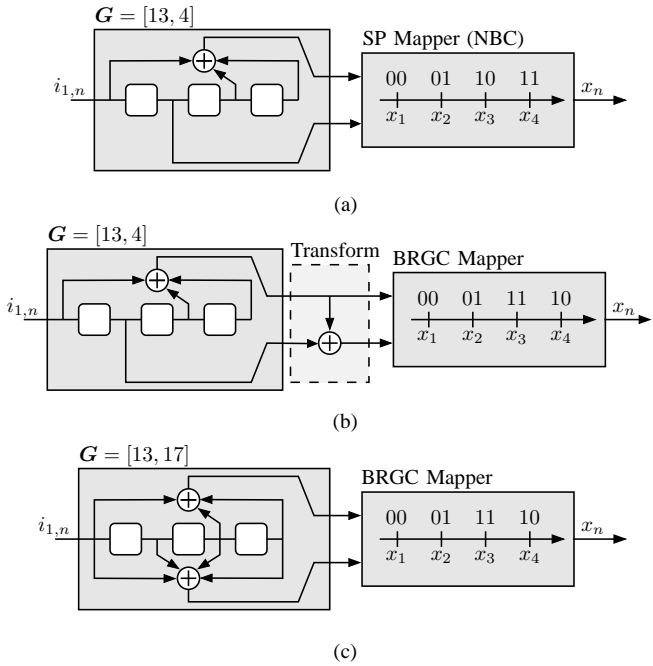}
    \caption{Three equivalent TCM encoders \cite{Fischer11PrivCommun}: (a) convolutional encoder with polynomial generators $\matG=[13,4]$ and an SP mapper \cite{Ungerboeck82}; (c) convolutional encoder with polynomial generators $\matG=[13,17]$ and a BRGC mapper \cite{Alvarado10d}. The encoder in (b) shows how a transformation based on a binary addition can be included in the mapper (to go from (b) to (a)) or in the encoder (to go from (b) to (c)).}
    \label{TCM_BICM-T_example}
\end{center}
\end{figure}

It was recently shown in \cite{Stierstorfer10} that if the interleaver is removed in BICM, its performance over the AWGN channel is greatly improved. This was later analyzed in detail in \cite{Alvarado10d} for a rate $R=1/2$ encoder and a $4$-ary pulse amplitude modulation (PAM) constellation, where the system in \cite{Stierstorfer10} was called ``BICM with trivial interleavers'' (BICM-T) and recognized as a TCM transmitter used with a BICM receiver. Moreover, BICM-T was shown to perform asymptotically as well as TCM (in terms of MED) \cite[Table~I]{Ungerboeck82} if properly chosen convolutional encoders are used \cite[Table~III]{Alvarado10d}. The transmitters in \cite[Table~I]{Ungerboeck82} and \cite[Table~III]{Alvarado10d} for the $8$-state (memory $\nu=3$) convolutional encoder\footnote{Throughout this paper, all polynomial generators are given in octal.} are shown in \figref{TCM_BICM-T_example}~(a) and \figref{TCM_BICM-T_example}~(c), respectively.

The authors in \cite{Alvarado10d} failed to note that in fact the optimal TCM encoder found when analyzing BICM-T is \emph{equivalent}\footnote{We use ``equivalent'' to denote two encoders with the same input-output relationship. This is formally defined in \secref{Sec:H}.} to the one proposed by Ungerboeck 30 years ago \cite{Fischer11PrivCommun}. For a $4$PAM constellation, one simple (although not unique) way of obtaining Ungerboeck's SP is by using the NBC. Moreover, the NBC can be generated using the BRGC plus one binary addition (which we call \emph{transform}) applied to its inputs, as shown in \figref{TCM_BICM-T_example}(b). If the transform is included in the mapper, the encoder in  \figref{TCM_BICM-T_example}(a) is obtained, while if it is included in the convolutional encoder, the TCM encoder in \figref{TCM_BICM-T_example}(c) is obtained. This equivalence also applies to encoders with larger number of states\footnote{This equivalence does not directly hold because \cite[Table~III]{Alvarado10d} lists the encoders in lexicographic order and because for some values of $\nu$ there are more than one encoder with identical performance.} and simply reveals that for $4$PAM, a TCM transceiver based on a BRGC mapper will have identical performance to Ungerboeck's TCM if the encoder is properly modified, where the modification is the application of a simple transform. The equivalence between TCM encoders and encoders optimized for the BRGC and the NBC as well as the relationship between the encoders in \cite{Alvarado10d} and \cite{Ungerboeck82} were first pointed out to us by R. F. H. Fischer \cite{Fischer11PrivCommun}. The idea of applying a linear transformation to the labeling/encoder can be traced back to\cite[Fig.~6.5]{Zhang96_Thesis} (see also \cite{Zhang94} and \cite[Ch.~2]{Gray99_Thesis}). 

TCM designs based on SP are considered heuristic \cite[Sec.~3.4]{Schlegel04_Book}, and thus, they do not necessarily lead to an optimal design \cite[p.~680]{Barry04_Book}.\footnote{Indeed, the results in \cite[Tables~2--3]{Du89}, \cite[Ch.~6]{Zhang96_Thesis} and \cite{Zhang94} show the suboptimality of the SP principle in terms of the multiplicities associated with the events at MED.} The problem of using non-SP labelings for TCM has been studied in \cite[Sec.~13.2.1]{Barry04_Book}, \cite[Sec.~8.6]{Clark81_Book}, and \cite{Viterbi89}. TCM encoders using the BRGC were designed in \cite{Du89}, by searching over convolutional encoders maximizing the MED. In \cite[Ch.~6]{Zhang96_Thesis} and \cite{Zhang94}, a non-Gray non-SP labeling was used and TCM encoders with optimal spectrum were tabulated.

In a related work, Wesel \emph{et al.} introduced in \cite{Wesel01} the concept of the edge profile (EP) of a labeling, and argued that in most cases, the EP can be used to find equivalent TCM encoders in terms of MED. The EP is also claimed to be a good indication of the quality of a labeling for TCM in \cite[Sec.~I]{Wesel01}; however, its optimality is not proven. Consequently, an exhaustive search over labelings with optimal EP does not necessarily lead to an optimal design \cite{Wesel12PrivCommun}.

In summary, as clearly explained in \cite[Sec.~I]{Wesel01}, traditional TCM designs either optimize the encoder for a constellation labeled using the SP principle, or simply connect a convolutional encoder designed for binary transmission with an ad-hoc binary labeling. It has been known for many years that optimal TCM encoders are obtained only by \emph{jointly} designing the convolutional encoder and the labeling of a TCM encoder \cite[p.~966]{Lin04_Book}. However, to the best of our knowledge, there are no works formally addressing this problem, and thus, optimal TCM encoders are yet to be found.

In this paper, we address the joint design of the feedforward convolutional encoder and the labeling for TCM. To this end, we show that binary labelings can be grouped into different classes that lead to equivalent TCM encoders. The classes are closely related to the \emph{Hadamard classes} introduced in \cite{Knagenhjelm96} in the context of vector quantization. This classification allows us to formally prove that in any TCM encoder, the NBC can be replaced by many other labelings (including the BRGC) without causing any performance degradation, provided that the encoder is properly selected. This explains the asymptotic equivalence between BICM-T and TCM observed in \cite{Alvarado10d}. Moreover, since the classification reduces the number of labelings that must be tested in an exhaustive search, we use it to tabulate optimal TCM encoders for $4$-ary and $8$-ary constellations.

\section{Preliminaries}\label{Sec:Prel}

\subsection{Notation Convention}\label{Sec:Prel.Notation}

Throughout this paper, scalars are denoted by italic letters $x$, row vectors by boldface letters $\bx=[x_{1},\ld,x_{N}]$, temporal sequences by underlined boldface letters $\un{\bx}=[\bx[1],\ld,\bx[\Ns]]$, and matrices by capital boldface letters $\matX$ where $x_{i,j}$ represents the entry of $\matX$ at row $i$, column $j$. The transpose of a matrix/vector is denoted by $[\cd]\T$. Matrices are sometimes expressed in the compact form $\matX=[\bx_1; \bx_2; \ld ; \bx_M]$, where $\bx_i=[x_{i,1}, \ld, x_{i,N}]$ is the $i$th row. Sets are denoted using calligraphic letters $\mc{C}$ and the binary set is defined as $\mcB\triangleq\set{0,1}$. Binary addition is denoted by $a \oplus b$.

The probability mass function (PMF) of the random variable $Y$ is denoted by $P_{Y}(y)$ and the probability density function (PDF) of the random variable $Y$ by $p_{Y}(y)$. Conditional PDFs are denoted as $p_{Y|X}(y|x)$. The tail probability of a standard Gaussian random variable is denoted by $Q(x) \triangleq \frac{1}{\sqrt{2\pi}}\int_{x}^{\infty}\tr{e}^{-{\xi^2}/{2}}\,\tr{d}\xi$.

\subsection{TCM Encoder}\label{Sec:Prel.Model}

\begin{figure}
\newcommand{\scale}{0.85}
\psfrag{bi1}[Br][Br][\scale]{$i_{1,n}$}
\psfrag{bik}[Br][Br][\scale]{$i_{k,n}$}
\psfrag{bb1}[Bl][Bl][\scale]{$u_{1,n}$}	
\psfrag{bbn}[Bl][Bl][\scale]{$u_{m,n}$}	
\psfrag{CONV}[cc][cc][\scale]{Conv.}
\psfrag{ENC}[cc][cc][\scale]{Encoder}
\psfrag{tcm}[Bl][Bl][\scale]{TCM Encoder}
\psfrag{ddd}[cc][cc][\scale][90]{$\cdots$}
\psfrag{MAP}[cc][cc][\scale]{$\Phi_\matL$}
\psfrag{bx}[Bl][Bl][\scale]{$\bx[n]$}
\begin{center}
	\includegraphics[width=0.6\columnwidth]{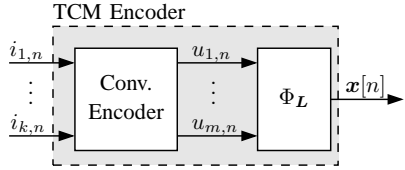}
	\caption{Generic TCM encoder under consideration: A feedforward convolutional encoder of rate $R=k/m$ with $2^\nu$ states serially concatenated with a memoryless $m$-bit mapper $\Phi_\matL$.}
	\label{tcm_encoder}
\end{center}
\end{figure}

We consider the TCM encoder shown in \figref{tcm_encoder} where a feedforward convolutional encoder of rate $R=k/m$ is serially connected to a mapper $\Phi_\matL$ and the index $\matL$ emphasizes the dependency of the mapper on the labeling (defined later). At each discrete time instant $n$, the information bits $i_{1,n},\ld,i_{k,n} $ are fed to the convolutional encoder, which is fully determined by $k$ different $\nu_p$-stage shift registers with $p=1,\ld,k$, and the way the input sequences are connected (through the registers) to its outputs. Closely following the notation of \cite[Sec.~11.1]{Lin04_Book}, we denote the \emph{memory} of the convolutional encoder by $\nu=\sum_{p=1}^{k}\nu_p$, and the \emph{number of states} by $2^\nu$. The connection between the input and output bits is defined by the binary representation of the \emph{convolutional encoder matrix} \cite[eq.~(11.6)]{Benedetto99_Book}
\begin{align}\label{Gmatrix}
\matG \triangleq
\left[
\begin{array}{cccc}
\bg_{1}^{(1)} 	& \bg_{1}^{(2)}	& \ld 	& \bg_{1}^{(m)}\\
\bg_{2}^{(1)} 	& \bg_{2}^{(2)}	& \ld 	& \bg_{2}^{(m)}\\
\vd 			& \vd		& \dd 	& \vd \\
\bg_{k}^{(1)} 	& \bg_{k}^{(2)}	& \ld 	& \bg_{k}^{(m)}\\
\end{array}
\right],
\end{align}
where $\bg_{p}^{(l)}\triangleq [g_{p,1}^{(l)},\ld,g_{p,\nu_p+1}^{(l)}]\T\in\mcB^{\nu_p+1}$ is a column vector representing the connection between the $p$th input sequence and the $l$th output sequence with $l=1,\ld,m$. The coefficients $g_{p,1}^{(l)},\ld,g_{p,\nu_p+1}^{(l)}$ are associated with the input bits $i_{p,n},\ld,i_{p,n-\nu_p}$, respectively, and $\matG\in\mcB^{(\nu+k)\times m}$. Throughout this paper, we will show the vectors $\bg_{p}^{(l)}$ defining $\matG$ either in binary or octal notation. When shown in octal notation, $g_{p,1}^{(l)}$ represents the most significant bit (see \figref{TCM_BICM-T_example}).

The convolutional encoder matrix \eqref{Gmatrix} allows us to express the output of the convolutional encoder at time $n$, which we define as $\bu_n\triangleq [u_{1,n},\ld,u_{m,n}]$, as a function of $(\nu+k)$ information bits, i.e.,
\begin{align}\label{IO}
\bu_n =  \bj_n \matG,
\end{align}
where $\bj_n\triangleq[\bi^{(1)}_n,\ld,\bi^{(k)}_n]$ with $\bi^{(p)}_n \triangleq [i_{p,n},\ld,i_{p,n-\nu_p}]$ are the information bits, and the matrix multiplication is in GF(2).

The coded bits $\bu_n$ are mapped to real $N$-dimensional constellation symbols using the mapper $\Phi_\matL:\mcB^m\rightarrow\mcX$, where $\mcX\subset\ms{R}^N$ is the constellation used for transmission, with $|\mcX|=M=2^m$. We use $\bx[n]\in\mcX$ to denote the transmitted symbols at time $n$ and the matrix $\matS=[\bx_1;\bx_2;\ld;\bx_M]$ with $\bx_q\in\ms{R}^N$ and $q=1,\ld,M$ to denote the ordered constellation points. We assume that the symbols are equally likely and that the constellation $\mc{X}$ is normalized to unit energy, i.e., $\Es\triangleq\ms{E}_{\matS}[\| \matS \|^2]=1/M\sum_{\bs\in\mc{X}}\|\bs\|^2=1$. As shown in \figref{tcm_encoder}, each symbol represents $k$ information bits.

The binary labeling of the $q$th symbol in $\matS$ is denoted by $\bc_q=[c_{q,1},\ld,c_{q,m}]\in\mcB^{m}$, where $c_{q,l}$ is the bit associated with the $l$th input of the mapper in \figref{tcm_encoder}. The labeling matrix is defined as $\matL=[\bc_{1};\bc_{2};\ld;\bc_{M}]$, where $\bc_q$ in $\matL$ corresponds to the binary label of the symbol $\bx_q$ in $\matS$. Throughout this paper, we will show the vectors $\bc_q$ in $\matL$ in either binary or integer notation.

\subsection{Binary Labelings for TCM}\label{Sec:Prel.Labelings}

The NBC of order $m$ is defined as $\matN_m\triangleq[\bn_1;\bn_2;\ld;\bn_M]$ where $\bn_q=[n_{q,1},\ld,n_{q,m}]\in\mcB^{m}$ is the base-2 representation of the integer $q-1$ and $n_{q,m}$ is the least significant bit. The BRGC of order $m$ is defined as $\matB_m\triangleq[\bb_1;\bb_2;\ld;\bb_M]$ where $\bb_q=[b_{q,1},\ld,b_{q,m}]\in\mcB^{m}$. The bits of the BRGC can be generated from the NBC as $b_{q,1}=n_{q,1}$ and $b_{q,l}=n_{q,l-1}\oplus n_{q,l}$ for $l=2,\ld,m$. Alternatively, we have $n_{q,l}=b_{q,1}\oplus \ld \oplus b_{q,l-1}\oplus b_{q,l}$ for $l=1,\ld,m$, or, in matrix notation, $\matB_m = \matN_m \matT$ and $\matN_m = \matB_m \matT^{-1}$, where
\begin{align}\label{T_BRGC_NBC}
\matT=
\left[
\begin{matrix}
1\,1\,0\,\ld\,0\,0\\
0\,1\,1\,\ld\,0\,0\\
0\,0\,1\,\ld\,0\,0\\
\vd\quad\dd\,\,\vd \\
0\,0\,0\,\ld\,1\,1\\
0\,0\,0\,\ld\,0\,1\\
\end{matrix}
\right],
\quad
\matT^{-1}=
\left[
\begin{matrix}
1\,1\,1\,\ld\,1\,1\\
0\,1\,1\,\ld\,1\,1\\
0\,0\,1\,\ld\,1\,1\\
\vd\quad\dd\,\,\vd \\
0\,0\,0\,\ld\,1\,1\\
0\,0\,0\,\ld\,0\,1\\
\end{matrix}
\right].
\end{align}

\begin{example}\label{Example:BRGC_NBC:m3}
The NBC and BRGC of order $m=3$ are
\begin{align}\label{Example:BRGC_NBC:m3.eq}
\matN_3=
\begin{bmatrix}
0\,0\,0 \\
0\,0\,{\bf 1} \\
0\,{\bf 1}\,0 \\
0\,1\,1 \\
{\bf 1}\,0\,0 \\
1\,0\,1 \\
1\,1\,0 \\
1\,1\,1 \\
\end{bmatrix}
,\quad
\matB_3=
\begin{bmatrix}
0\,0\,0 \\
0\,0\,{\bf 1} \\
0\,{\bf 1}\,1 \\
0\,1\,0 \\
{\bf 1}\,1\,0 \\
1\,1\,1 \\
1\,0\,1 \\
1\,0\,0 \\
\end{bmatrix},
\end{align}
where the pivots of the labeling matrices (defined in \secref{Sec:H.Preliminaries}) are highlighted.
\end{example}

To formally define the SP principle for a given constellation $\matS$ and labeling $\matL$, we define $\mc{X}_l([u_{m+1-l},\ld,u_m])\triangleq\set{\bx_q\in\mcX:[c_{q,m+1-l},\ld,c_{q,m}]=[u_{m+1-l},\ld,u_m], q=1,\ld,M}\subset\mc{X}$ for $l=1,\ld,m-1$. Additionally, we define the minimum intra-Euclidean distance (intra-ED) at level $l$ as
\begin{align}
\delta_{l} \triangleq \min_{\substack{\bx_i,\bx_j\in \mc{X}_l(\bu)\\ i\neq j, \bu\in\mcB^{l}}} \|\bx_i-\bx_j\|, \quad l=1,\ld,m-1.
\end{align}
and the MED of the constellation as $\delta_{0}$.

\begin{definition}[Set-partitioning \cite{Ungerboeck82}]
\label{SP.definition}
For a given constellation $\matS$, the labeling $\matL$ is said to follow the SP principle if $\delta_{0} < \delta_{1} < \ld < \delta_{m-1}$.
\end{definition}

%

\begin{example}
\label{SP-labelings-8psk}
Consider an $8$PSK constellation (formally defined in \secref{Sec:Results}). It can be easily verified that if this constellation is labeled by the NBC in \eqref{Example:BRGC_NBC:m3.eq}, an SP-labeled constellation is obtained. Although the NBC is the most intuitive form for generating an SP labeling for $M$PSK constellations, it is not unique. As an example, consider the semi set-partitioning (SSP) labeling proposed in \cite[Fig.~2(c)]{Li02} and the so-called modified set-partitioning (MSP) labeling \cite[Fig.~2(b)]{Tran06a}:
\begin{align}\label{Example:SSP_MSP:m3.eq}
\matL_{\tr{SSP}}=
\begin{bmatrix}
0\,0\,0 \\
1\,0\,1 \\
0\,1\,0 \\
1\,1\,1 \\
1\,0\,0 \\
0\,0\,1 \\
1\,1\,0 \\
0\,1\,1 \\
\end{bmatrix}
,\quad
\matL_{\tr{MSP}}=
\begin{bmatrix}
0\,0\,0 \\
0\,0\,1 \\
0\,1\,0 \\
1\,1\,1 \\
1\,0\,0 \\
1\,0\,1 \\
1\,1\,0 \\
0\,1\,1 \\
\end{bmatrix}.
\end{align}
 It can be shown that both labelings follow the SP principle in Definition~\ref{SP.definition}.
 \end{example}

Example~\ref{SP-labelings-8psk} shows that there are multiple labelings that follow the SP principle. It can be shown that this is also the case for $M$PAM constellations, and that in this case, the NBC is also an SP labeling.

\subsection{System Optimization and Search Problems}\label{Sec:Prel.Optimization}

For a given constellation $\matS$ and memory $\nu$, a \emph{TCM encoder} is fully defined by the convolutional encoder matrix $\matG$ and the labeling of the constellation $\matL$, and thus, a TCM encoder is defined by the pair $\Theta=[\matG,\matL]$.

For given integers $k$, $m$, and $\nu$, we define the \emph{convolutional encoder universe} as the set $\mc{G}_{k,m,\nu}$ of all $(\nu +k)\times m$ binary matrices\footnote{Note that whenever $\matG$ is given in its binary form, $\nu_1,\dots,\nu_k$ are also needed to interpret $\matG$ correctly according to \eqref{Gmatrix}.} $\matG$ which result in a noncatastrophic feedforward encoder and equally likely symbols.\footnote{For some matrices $\matG$, the symbols $\bx[n]$ can be nonequally likely. This would induce nonequally likely symbols (signal shaping) which we do not consider in this work.} We are also interested in the \emph{labeling universe}, defined for a given integer $m$ as the set $\mc{L}_m$ of all $M\times m$ binary matrices whose $M$ rows are all distinct.

To the best of our knowledge, there are no works addressing the problem of designing a TCM encoder by \emph{exhaustively} searching over the labeling universe and the convolutional encoder universe. We believe the reason for this is that an exhaustive search over encoders and labelings is unfeasible \cite[Sec.~I]{Benedetto88}. For example, for $8$-ary constellations, there are in general $8!=40320$ different binary labelings. In this paper, we show how a joint optimization over all $\matG \in \mc{G}_{k,m,\nu}$ and $\matL \in \mc{L}_m$ can be restricted, without loss of generality, to a joint optimization over all $\matG \in \mc{G}_{k,m,\nu}$ and a subset of $\mc{L}_m$.

\section{Equivalent Labelings for TCM Encoders}\label{Sec:H}

In this section, we show that binary labelings can be grouped into classes, and that all the labelings belonging to the same class lead to equivalent TCM encoders. This analysis is inspired by the one in \cite{Knagenhjelm96}, where the so-called Hadamard classes were used to solve a related search problem in source coding.

\subsection{Equivalent TCM Encoders}\label{Sec:H.Equivalences}

The transmitted symbol at time $n$ of a given TCM encoder $\Theta=[\matG,\matL]$ can be expressed using \eqref{IO} as
\begin{align}\label{IO.symbol}
\bx[n]=\Phi_{\matL}(\bu_n) = \Phi_{\matL}(\bj_n \matG).
\end{align}

\begin{definition}\label{TCM.Equivalence}
Two TCM encoders $\Theta=[\matG,\matL]$ and $\tilde{\Theta}=[\tilde{\matG},\tilde{\matL}]$ are said to be \emph{equivalent} if they give the same output symbol for the same information bit sequence, \ie if they fulfill
$\Phi_{\matL}(\bj \matG)=\Phi_{\tilde{\matL}}(\bj \tilde{\matG})$
for any $\bj\in\mcB^{\nu+k}$.
\end{definition}

The concept of ``equivalent encoders'' is more restrictive than the more well-known concept of ``equivalent codes''. Two equivalent encoders have the same bit error rate (BER) and frame error rate (FER), whereas two equivalent codes have the same FER but in general different BER. In this paper, where BER is an important figure of merit, we are therefore more interested in equivalent encoders.

%

From now on we use $\mcT_m$ to denote the set of all binary invertible $m\times m$ matrices.

\begin{lemma}\label{lemma.mapper}
$\Phi_{\matL}(\bc)=\Phi_{\tilde{\matL}}(\bc\matT)$ where $\tilde{\matL} = \matL \matT$,  for any two mappers $\Phi_{\matL}$ and $\Phi_{\tilde{\matL}}$ that use the same constellation $\matX$, any $\matT \in \mcT_m$, and any $\bc\in\mcB^m$.
\end{lemma}

\begin{IEEEproof}
Let ${\bv}_q \triangleq [0,\ld,0,1,0,\ld,0]$ be a vector of length $M$, where the one is in position $q$. From the definition of the labeling matrix $\matL$, it follows that $\bc_q = {\bv}_q \matL$ for $q=1,\ld,M$. The mapping $\Phi_{\matL}$ satisfies by definition $\Phi_{\matL}(\bc_q) = \bx_q$ for $q=1,\ld,M$, or, making the dependency on $\matL$ explicit,
\begin{align}\label{eq:phi1}
\Phi_{\matL}(\bc) = \bx_q, \quad \text{if } \bc = {\bv}_q \matL
\end{align}
for any $\bc \in \mcB^m$. Similarly, for any $\bc \in \mcB^m$,
\begin{align}\label{eq:phi2}
\Phi_{\tilde{\matL}}(\bc\matT) &= \bx_q, \quad \text{if } \bc\matT = {\bv}_q \tilde{\matL} \nonumber\\
&= \bx_q, \quad \text{if } \bc = {\bv}_q \matL,
\end{align}
where the last step follows because $\matL = \tilde{\matL} \matT^{-1}$. Since the right-hand sides of \eqref{eq:phi1} and \eqref{eq:phi2} are equal, $\Phi_{\tilde{\matL}}(\bc\matT) = \Phi_{\matL}(\bc)$ for all $\bc \in \mcB^m$.
\end{IEEEproof}

The following theorem is the main result of this paper.
\begin{theorem}\label{theo.equivalence}
For any $\matG \in \mc{G}_{k,m,\nu}$, $\matL\in\mc{L}_m$, and $\matT\in\mcT_m$, the two TCM encoders $\Theta=[\matG,\matL]$ and $\tilde{\Theta}=[\tilde{\matG},\tilde{\matL}]$ are equivalent, where $\tilde{\matL}=\matL \matT$ and $\tilde{\matG}=\matG\matT$.
\end{theorem}

\begin{IEEEproof}
For any $\bj\in\mcB^{\nu+k}$, $\Phi_{\tilde{\matL}}(\bj \tilde{\matG}) = \Phi_{\tilde{\matL}}(\bj \matG\matT) = \Phi_{\matL}(\bj \matG)$, where the last equality follows by Lemma~\ref{lemma.mapper}. The theorem now follows using Definition~\ref{TCM.Equivalence}.
\end{IEEEproof}

Theorem~\ref{theo.equivalence} shows that a full search over $\mc{G}_{k,m,\nu}$ and $\mc{L}_m$ will include many pairs of equivalent TCM encoders. Therefore, an optimal TCM encoder with given parameters can be found by searching over a subset of $\mc{G}_{k,m,\nu}$ and  the whole set $\mc{L}_m$ or vice versa. In this paper, we choose the latter approach, searching over a subset of $\mc{L}_m$.

\subsection{Matrix Factorization}\label{Sec:H.Preliminaries}

We briefly summarize here some matrix algebra. The following definition of a \emph{reduced column echelon matrix} comes from \cite[pp.~183--184]{Birkhoff77_Book}, adapted to the fact that we only consider binary labeling matrices $\matL$ whose columns are all nonzero. The first nonzero element of the $k$th column is called the $k$th \emph{pivot} of $\matL$. The pivots for $\matN_3$ and $\matB_3$ are highlighted in \eqref{Example:BRGC_NBC:m3.eq}.

\begin{definition}
\label{Def.REM}
A matrix $\matL\in\mcB^{M\times m}$ is a reduced column echelon matrix if the following two conditions are fulfilled:
\begin{enumerate}
\item Every row with a pivot has all its other entries zero.
\item The pivot in column $l$ is located in a row below the pivot in column $l+1$, for $l=1,\ld,m-1$.
\end{enumerate}
\end{definition}

The matrix $\matN_3$ in Example~\ref{Example:BRGC_NBC:m3} (or more generally $\matN_m$) is an example of a reduced column echelon matrix. On the other hand, $\matB_m$ is not a reduced column echelon matrix because it does not fulfill the first condition in Definition~\ref{Def.REM}.

The following theorem will be used to develop an efficient search algorithm in the next section. We refer the reader to \cite[p.~187, Corollary~1]{Birkhoff77_Book} for a proof. From now on we use $\mcR_m$ to denote the set of all reduced column echelon binary matrices.
\begin{theorem}\label{LT.Theorem}
Any binary labeling $\matL\in\mc{L}_m$ can be uniquely factorized as
\begin{align}\label{LT}
\matL = \matL_\tr{R}\matT,
\end{align}
where $\matT\in\mcT_m$ and $\matL_\tr{R}\in\mcR_m$.
\end{theorem}

Theorem~\ref{LT.Theorem} shows that all binary labeling matrices $\matL$ can be uniquely generated by finding all the invertible matrices $\matT$ (the set $\mcT_m$) and all reduced column echelon matrices $\matL_\tr{R}$ (the set $\mcR_m$). In particular, we have \cite[eq.~(1)]{Duvall71}, \cite[eq.~(18)]{Knagenhjelm96}
\begin{gather}
\label{Mi}
M_\tr{T} \triangleq |\mcT_m| = \prod_{l=1}^{m}(2^m-2^{l-1}), \\
\label{Me}
M_\tr{R} \triangleq |\mcR_m| = \frac{2^m!}{\prod_{l=1}^{m}(2^m-2^{l-1})}.
\end{gather}
In \tabref{Table_Mi_Me}, the values for $M_\tr{R}$ and $M_\tr{T}$ for $1\leq m \leq 6$ are shown. In this table we also show the number of binary labelings ($|\mc{L}_m|=2^m!=M_\tr{R}M_\tr{T}$), i.e., the number of matrices $\matL$ in the labeling universe.

\begin{table}
\small\centering
\renewcommand{\arraystretch}{1.4}
\caption{Number of classes ($M_\tr{R}=|\mcR_m|$), their cardinality ($M_\tr{T}=|\mcT_m|$), and the total number of labelings ($2^m!$) for different values of $m$.}
\begin{center}
\begin{tabular}{@{~~}c@{~~}c@{~~}c@{~~}c@{~~}c@{~~}c@{~~}c@{~~}}
\hline

\hline
$m$			& $1$ & $2$& $3$ 	& $4$ 				& $5$ 				& $6$\\
\hline

\hline
	$M_\tr{R}$ & 2 	& 4 	& 240 	& $1.038\cd10^{9}$		& $2.632\cd 10^{28}$	& $6.294\cd 10^{78}$\\
	$M_\tr{T}$  & 1 	& 6 	& 168 	& 20160 				& $9.999\cd 10^{6}$		& $2.016\cd 10^{10}$\\
	$2^m!$		& 2 	& 24 & 40320	& $2.092\cd10^{13}$	& $2.631\cd 10^{35}$ 	& $1.269\cd 10^{89}$\\
\hline

\hline
\end{tabular}
\end{center}
\label{Table_Mi_Me}
\end{table}

The \emph{modified Hadamard class} associated with the reduced column echelon matrix $\matL_\tr{R}$ is defined as the set of matrices $\matL$ that can be generated via \eqref{LT} by applying all $\matT\in\mcT_m$. Note that these modified Hadamard classes are narrower than the regular Hadamard classes defined in \cite{Knagenhjelm96}, each including $M$ reduced column echelon matrices. There are thus $M_\tr{R}$ modified Hadamard classes, each with cardinality $M_\tr{T}$.

As a consequence of Theorems~\ref{theo.equivalence} and \ref{LT.Theorem}, the two TCM encoders $[\matG,\matL]$ and $[\matG\matT^{-1},\matL_\tr{R}]$ are equivalent for any $\matG \in \mc{G}_{k,m,\nu}$ and $\matL\in\mc{L}_m$, where $\matL_\tr{R}$ and $\matT$ are given by the factorization \eqref{LT}. In other words, all nonequivalent TCM encoders can be generated using one member of each modified Hadamard class only, and thus, a joint optimization over all $\matG \in \mc{G}_{k,m,\nu}$ and $\matL \in \mc{L}_m$ can be reduced to an optimization over all $\matG \in \mc{G}_{k,m,\nu}$ and $\matL \in \mcR_m$ with no loss in performance. This means that the search space is reduced by at least a factor of $M_\tr{T}=M!/M_\tr{R}$. For example, for $8$-ary constellations ($m=3$), the total number of different binary labelings that must be tested is reduced from $8!=40320$ to $240$. Moreover, as we will see in \secref{Sec:Results}, this can be reduced even further if the constellation $\matX$ possesses certain symmetries.

\subsection{Modified Full Linear Search Algorithm}\label{Sec:HMFSLA}

The problem of finding the set $\mcR_m$ of reduced column echelon matrices for a given $m$ can be solved by using a modified version of the full linear search algorithm (FLSA) introduced in \cite[Sec.~VIII]{Knagenhjelm96}. We call this algorithm the modified FLSA (MFLSA). The MFLSA generates one member of each modified Hadamard class, the one that corresponds to a reduced column echelon matrix $\matL_\tr{R}$. Its pseudocode is shown in Algorithm~1. In this algorithm, the vector $\br=[r_1,\ld,r_{M}]$ denotes the integer representation of the rows of the matrix $\matL_\tr{R}$ where $r_q=c_{q,m}+2c_{q,m-1}+\ld+2^{m-1}c_{q,1}$ for $q=1,\ld,M$. The first labeling generated (line 1) is always the NBC. Then the algorithm proceeds by generating all permutations thereof, under the condition that no power of two ($1,2,4,\ld$) is preceded by a larger value. By Definition~\ref{Def.REM}, this simple condition assures that only reduced column echelon matrices are generated.

\renewcommand{\algorithmicrequire}{\textbf{Input:}}
\renewcommand{\algorithmicensure}{\textbf{Output:}}
\begin{algorithm}[t]
{\small
	\caption{Modified full linear search algorithm (MFLSA)}
	\label{alg:MFLSA}
	\begin{algorithmic}[1]
		\REQUIRE The order $m$
		\ENSURE Print the $M_\tr{R}$ different reduced column echelon vectors $\br$
		\STATE $\br \leftarrow [0, 1, \ld , M-1]$
		\LOOP
			\PRINT $\br$
			\STATE $\ind \leftarrow 0$
			\WHILE{$r_{M}=\ind$}
				\STATE $[r_{\ind+1},\ld,r_{M}] \leftarrow [r_{M}, r_{\ind+1},\ld,r_{M-1}]$
				\STATE $\ind \leftarrow \ind+1$
				\WHILE{$\ind$ is a power of 2}
					\STATE $\ind \leftarrow \ind+1$
				\ENDWHILE
				\IF{$\ind = M-1$}
					\STATE Quit
				\ENDIF
			\ENDWHILE
			\STATE Find $\pointer$ such that $r_{\pointer}=\ind$
			\STATE Swap $r_{\pointer}$ and $r_{\pointer+1}$
		\ENDLOOP
	\end{algorithmic}
}
\end{algorithm}

\begin{example}
For $m=2$, the MFLSA returns the following reduced column echelon matrices:
\begin{align}\label{MFLSA.m.2}
\mcR_2 &=
\left\{
\begin{bmatrix}
0\,0 \\
0\,{\bf 1} \\
{\bf 1}\,0 \\
1\,1 \\
\end{bmatrix}
,
\begin{bmatrix}
0\,{\bf 1} \\
0\,0 \\
{\bf 1}\,0 \\
1\,1 \\
\end{bmatrix}
,
\begin{bmatrix}
0\,{\bf 1} \\
{\bf 1}\,0 \\
0\,0 \\
1\,1 \\
\end{bmatrix}
,
\begin{bmatrix}
0\,{\bf 1} \\
{\bf 1}\,0 \\
1\,1 \\
0\,0 \\
\end{bmatrix}
\right\},
\end{align}
where the first element in $\mcR_2$ is the NBC defined in \secref{Sec:Prel.Labelings} and again we highlighted the pivots of the matrices. The 6 binary invertible matrices for $m=2$ are
\begin{align}\label{T.m.2}
\mcT_2
=\left\{
\begin{bmatrix}
0\,1\\
1\,0\\
\end{bmatrix},
\begin{bmatrix}
0\,1\\
1\,1\\
\end{bmatrix},
\begin{bmatrix}
1\,0\\
0\,1\\
\end{bmatrix},
\begin{bmatrix}
1\,0\\
1\,1\\
\end{bmatrix},
 \begin{bmatrix}
1\,1\\
0\,1\\
\end{bmatrix},
\begin{bmatrix}
1\,1\\
1\,0\\
\end{bmatrix}
\right\}.
\end{align}
Using Theorem~\ref{LT.Theorem}, all the 24 binary labelings in $\mc{L}_2$ (see \tabref{Table_Mi_Me}) can be generated by multiplying the matrices in $\mcR_2$ and $\mcT_2$.
\end{example}

\begin{example}
For $m=3$, the reduced column echelon matrices generated by the MFLSA are shown in \tabref{Table.MFLSA.m3} (in integer notation). The MFLSA first generates row number one, then row number two, then row number three, etc., where each row is generated from left to right. The first column in the table corresponds to the output of the FLSA of \cite{Knagenhjelm96}. Columns two to eight show the additional matrices generated by the MFLSA, which are obtained from the first column by shifting the symbol zero to the right. In this table we also highlight the labelings generated by the MFLSA that at the same time have optimal EP \cite{Wesel01} for $8$PAM and $8$PSK (see \secref{Sec:Results}).

\begin{table*}
\caption{Reduced column echelon matrices for $m=3$ generated by the MFLSA. The MFLSA first generates row number one, then row number two, etc. The labelings shown in boldface have optimal EP for 8PAM (first four columns) and for 8PSK (first column).}
\label{Table.MFLSA.m3}
\scriptsize
\begin{center}
\begin{tabular}{cccccccc}
\LOEP{0\,1\,2\,3\,4\,5\,6\,7} & 1\,0\,2\,3\,4\,5\,6\,7 & 1\,2\,0\,3\,4\,5\,6\,7 & 1\,2\,3\,0\,4\,5\,6\,7 & 1\,2\,3\,4\,0\,5\,6\,7 & 1\,2\,3\,4\,5\,0\,6\,7 & 1\,2\,3\,4\,5\,6\,0\,7 & {1\,2\,3\,4\,5\,6\,7\,0} \\ \hline
0\,1\,2\,4\,3\,5\,6\,7 & 1\,0\,2\,4\,3\,5\,6\,7 & 1\,2\,0\,4\,3\,5\,6\,7 & 1\,2\,4\,0\,3\,5\,6\,7 & 1\,2\,4\,3\,0\,5\,6\,7 & 1\,2\,4\,3\,5\,0\,6\,7 & {1\,2\,4\,3\,5\,6\,0\,7} & 1\,2\,4\,3\,5\,6\,7\,0 \\ \hline
0\,1\,2\,4\,5\,3\,6\,7 & 1\,0\,2\,4\,5\,3\,6\,7 & 1\,2\,0\,4\,5\,3\,6\,7 & 1\,2\,4\,0\,5\,3\,6\,7 & 1\,2\,4\,5\,0\,3\,6\,7 & {1\,2\,4\,5\,3\,0\,6\,7} & 1\,2\,4\,5\,3\,6\,0\,7 & 1\,2\,4\,5\,3\,6\,7\,0 \\ \hline
0\,1\,2\,4\,5\,6\,3\,7 & 1\,0\,2\,4\,5\,6\,3\,7 & 1\,2\,0\,4\,5\,6\,3\,7 & 1\,2\,4\,0\,5\,6\,3\,7 & 1\,2\,4\,5\,0\,6\,3\,7 & 1\,2\,4\,5\,6\,0\,3\,7 & 1\,2\,4\,5\,6\,3\,0\,7 & 1\,2\,4\,5\,6\,3\,7\,0 \\ \hline
0\,1\,2\,4\,5\,6\,7\,3 & 1\,0\,2\,4\,5\,6\,7\,3 & 1\,2\,0\,4\,5\,6\,7\,3 & 1\,2\,4\,0\,5\,6\,7\,3 & 1\,2\,4\,5\,0\,6\,7\,3 & 1\,2\,4\,5\,6\,0\,7\,3 & 1\,2\,4\,5\,6\,7\,0\,3 & 1\,2\,4\,5\,6\,7\,3\,0 \\ \hline
0\,1\,2\,3\,4\,6\,5\,7 & 1\,0\,2\,3\,4\,6\,5\,7 & 1\,2\,0\,3\,4\,6\,5\,7 & 1\,2\,3\,0\,4\,6\,5\,7 & 1\,2\,3\,4\,0\,6\,5\,7 & 1\,2\,3\,4\,6\,0\,5\,7 & 1\,2\,3\,4\,6\,5\,0\,7 & 1\,2\,3\,4\,6\,5\,7\,0 \\ \hline
0\,1\,2\,4\,3\,6\,5\,7 & 1\,0\,2\,4\,3\,6\,5\,7 & 1\,2\,0\,4\,3\,6\,5\,7 & 1\,2\,4\,0\,3\,6\,5\,7 & 1\,2\,4\,3\,0\,6\,5\,7 & 1\,2\,4\,3\,6\,0\,5\,7 & 1\,2\,4\,3\,6\,5\,0\,7 & 1\,2\,4\,3\,6\,5\,7\,0 \\ \hline
0\,1\,2\,4\,6\,3\,5\,7 & 1\,0\,2\,4\,6\,3\,5\,7 & 1\,2\,0\,4\,6\,3\,5\,7 & 1\,2\,4\,0\,6\,3\,5\,7 & {1\,2\,4\,6\,0\,3\,5\,7} & 1\,2\,4\,6\,3\,0\,5\,7 & 1\,2\,4\,6\,3\,5\,0\,7 & 1\,2\,4\,6\,3\,5\,7\,0 \\ \hline
0\,1\,2\,4\,6\,5\,3\,7 & 1\,0\,2\,4\,6\,5\,3\,7 & 1\,2\,0\,4\,6\,5\,3\,7 & \LOEP{1\,2\,4\,0\,6\,5\,3\,7} & 1\,2\,4\,6\,0\,5\,3\,7 & 1\,2\,4\,6\,5\,0\,3\,7 & 1\,2\,4\,6\,5\,3\,0\,7 & 1\,2\,4\,6\,5\,3\,7\,0 \\ \hline
0\,1\,2\,4\,6\,5\,7\,3 & 1\,0\,2\,4\,6\,5\,7\,3 & \LOEP{1\,2\,0\,4\,6\,5\,7\,3} & 1\,2\,4\,0\,6\,5\,7\,3 & 1\,2\,4\,6\,0\,5\,7\,3 & 1\,2\,4\,6\,5\,0\,7\,3 & 1\,2\,4\,6\,5\,7\,0\,3 & 1\,2\,4\,6\,5\,7\,3\,0 \\ \hline
0\,1\,2\,3\,4\,6\,7\,5 & 1\,0\,2\,3\,4\,6\,7\,5 & 1\,2\,0\,3\,4\,6\,7\,5 & 1\,2\,3\,0\,4\,6\,7\,5 & 1\,2\,3\,4\,0\,6\,7\,5 & 1\,2\,3\,4\,6\,0\,7\,5 & 1\,2\,3\,4\,6\,7\,0\,5 & 1\,2\,3\,4\,6\,7\,5\,0 \\ \hline
0\,1\,2\,4\,3\,6\,7\,5 & 1\,0\,2\,4\,3\,6\,7\,5 & 1\,2\,0\,4\,3\,6\,7\,5 & 1\,2\,4\,0\,3\,6\,7\,5 & 1\,2\,4\,3\,0\,6\,7\,5 & 1\,2\,4\,3\,6\,0\,7\,5 & 1\,2\,4\,3\,6\,7\,0\,5 & 1\,2\,4\,3\,6\,7\,5\,0 \\ \hline
0\,1\,2\,4\,6\,3\,7\,5 & 1\,0\,2\,4\,6\,3\,7\,5 & 1\,2\,0\,4\,6\,3\,7\,5 & 1\,2\,4\,0\,6\,3\,7\,5 & 1\,2\,4\,6\,0\,3\,7\,5 & 1\,2\,4\,6\,3\,0\,7\,5 & 1\,2\,4\,6\,3\,7\,0\,5 & 1\,2\,4\,6\,3\,7\,5\,0 \\ \hline
0\,1\,2\,4\,6\,7\,3\,5 & 1\,0\,2\,4\,6\,7\,3\,5 & 1\,2\,0\,4\,6\,7\,3\,5 & 1\,2\,4\,0\,6\,7\,3\,5 & 1\,2\,4\,6\,0\,7\,3\,5 & 1\,2\,4\,6\,7\,0\,3\,5 & 1\,2\,4\,6\,7\,3\,0\,5 & 1\,2\,4\,6\,7\,3\,5\,0 \\ \hline
0\,1\,2\,4\,6\,7\,5\,3 & \LOEP{1\,0\,2\,4\,6\,7\,5\,3} & 1\,2\,0\,4\,6\,7\,5\,3 & 1\,2\,4\,0\,6\,7\,5\,3 & 1\,2\,4\,6\,0\,7\,5\,3 & 1\,2\,4\,6\,7\,0\,5\,3 & 1\,2\,4\,6\,7\,5\,0\,3 & 1\,2\,4\,6\,7\,5\,3\,0 \\ \hline
0\,1\,2\,3\,4\,5\,7\,6 & \LOEP{1\,0\,2\,3\,4\,5\,7\,6} & 1\,2\,0\,3\,4\,5\,7\,6 & 1\,2\,3\,0\,4\,5\,7\,6 & 1\,2\,3\,4\,0\,5\,7\,6 & 1\,2\,3\,4\,5\,0\,7\,6 & 1\,2\,3\,4\,5\,7\,0\,6 & 1\,2\,3\,4\,5\,7\,6\,0 \\ \hline
0\,1\,2\,4\,3\,5\,7\,6 & 1\,0\,2\,4\,3\,5\,7\,6 & 1\,2\,0\,4\,3\,5\,7\,6 & 1\,2\,4\,0\,3\,5\,7\,6 & 1\,2\,4\,3\,0\,5\,7\,6 & 1\,2\,4\,3\,5\,0\,7\,6 & 1\,2\,4\,3\,5\,7\,0\,6 & 1\,2\,4\,3\,5\,7\,6\,0 \\ \hline
0\,1\,2\,4\,5\,3\,7\,6 & 1\,0\,2\,4\,5\,3\,7\,6 & 1\,2\,0\,4\,5\,3\,7\,6 & 1\,2\,4\,0\,5\,3\,7\,6 & 1\,2\,4\,5\,0\,3\,7\,6 & 1\,2\,4\,5\,3\,0\,7\,6 & 1\,2\,4\,5\,3\,7\,0\,6 & 1\,2\,4\,5\,3\,7\,6\,0 \\ \hline
0\,1\,2\,4\,5\,7\,3\,6 & 1\,0\,2\,4\,5\,7\,3\,6 & 1\,2\,0\,4\,5\,7\,3\,6 & 1\,2\,4\,0\,5\,7\,3\,6 & 1\,2\,4\,5\,0\,7\,3\,6 & 1\,2\,4\,5\,7\,0\,3\,6 & 1\,2\,4\,5\,7\,3\,0\,6 & 1\,2\,4\,5\,7\,3\,6\,0 \\ \hline
0\,1\,2\,4\,5\,7\,6\,3 & 1\,0\,2\,4\,5\,7\,6\,3 & 1\,2\,0\,4\,5\,7\,6\,3 & 1\,2\,4\,0\,5\,7\,6\,3 & 1\,2\,4\,5\,0\,7\,6\,3 & 1\,2\,4\,5\,7\,0\,6\,3 & 1\,2\,4\,5\,7\,6\,0\,3 & 1\,2\,4\,5\,7\,6\,3\,0 \\ \hline
0\,1\,2\,3\,4\,7\,5\,6 & 1\,0\,2\,3\,4\,7\,5\,6 & \LOEP{1\,2\,0\,3\,4\,7\,5\,6} & 1\,2\,3\,0\,4\,7\,5\,6 & 1\,2\,3\,4\,0\,7\,5\,6 & 1\,2\,3\,4\,7\,0\,5\,6 & 1\,2\,3\,4\,7\,5\,0\,6 & 1\,2\,3\,4\,7\,5\,6\,0 \\ \hline
0\,1\,2\,4\,3\,7\,5\,6 & 1\,0\,2\,4\,3\,7\,5\,6 & 1\,2\,0\,4\,3\,7\,5\,6 & 1\,2\,4\,0\,3\,7\,5\,6 & 1\,2\,4\,3\,0\,7\,5\,6 & 1\,2\,4\,3\,7\,0\,5\,6 & 1\,2\,4\,3\,7\,5\,0\,6 & 1\,2\,4\,3\,7\,5\,6\,0 \\ \hline
0\,1\,2\,4\,7\,3\,5\,6 & 1\,0\,2\,4\,7\,3\,5\,6 & 1\,2\,0\,4\,7\,3\,5\,6 & 1\,2\,4\,0\,7\,3\,5\,6 & {1\,2\,4\,7\,0\,3\,5\,6} & 1\,2\,4\,7\,3\,0\,5\,6 & 1\,2\,4\,7\,3\,5\,0\,6 & 1\,2\,4\,7\,3\,5\,6\,0 \\ \hline
0\,1\,2\,4\,7\,5\,3\,6 & 1\,0\,2\,4\,7\,5\,3\,6 & 1\,2\,0\,4\,7\,5\,3\,6 & 1\,2\,4\,0\,7\,5\,3\,6 & 1\,2\,4\,7\,0\,5\,3\,6 & 1\,2\,4\,7\,5\,0\,3\,6 & 1\,2\,4\,7\,5\,3\,0\,6 & 1\,2\,4\,7\,5\,3\,6\,0 \\ \hline
0\,1\,2\,4\,7\,5\,6\,3 & 1\,0\,2\,4\,7\,5\,6\,3 & 1\,2\,0\,4\,7\,5\,6\,3 & 1\,2\,4\,0\,7\,5\,6\,3 & 1\,2\,4\,7\,0\,5\,6\,3 & 1\,2\,4\,7\,5\,0\,6\,3 & {1\,2\,4\,7\,5\,6\,0\,3} & 1\,2\,4\,7\,5\,6\,3\,0 \\ \hline
0\,1\,2\,3\,4\,7\,6\,5 & 1\,0\,2\,3\,4\,7\,6\,5 & 1\,2\,0\,3\,4\,7\,6\,5 & \LOEP{1\,2\,3\,0\,4\,7\,6\,5} & 1\,2\,3\,4\,0\,7\,6\,5 & 1\,2\,3\,4\,7\,0\,6\,5 & 1\,2\,3\,4\,7\,6\,0\,5 & 1\,2\,3\,4\,7\,6\,5\,0 \\ \hline
0\,1\,2\,4\,3\,7\,6\,5 & 1\,0\,2\,4\,3\,7\,6\,5 & 1\,2\,0\,4\,3\,7\,6\,5 & 1\,2\,4\,0\,3\,7\,6\,5 & 1\,2\,4\,3\,0\,7\,6\,5 & 1\,2\,4\,3\,7\,0\,6\,5 & 1\,2\,4\,3\,7\,6\,0\,5 & 1\,2\,4\,3\,7\,6\,5\,0 \\ \hline
0\,1\,2\,4\,7\,3\,6\,5 & 1\,0\,2\,4\,7\,3\,6\,5 & 1\,2\,0\,4\,7\,3\,6\,5 & 1\,2\,4\,0\,7\,3\,6\,5 & 1\,2\,4\,7\,0\,3\,6\,5 & {1\,2\,4\,7\,3\,0\,6\,5} & 1\,2\,4\,7\,3\,6\,0\,5 & 1\,2\,4\,7\,3\,6\,5\,0 \\ \hline
0\,1\,2\,4\,7\,6\,3\,5 & 1\,0\,2\,4\,7\,6\,3\,5 & 1\,2\,0\,4\,7\,6\,3\,5 & 1\,2\,4\,0\,7\,6\,3\,5 & 1\,2\,4\,7\,0\,6\,3\,5 & 1\,2\,4\,7\,6\,0\,3\,5 & 1\,2\,4\,7\,6\,3\,0\,5 & 1\,2\,4\,7\,6\,3\,5\,0 \\ \hline
\LOEP{0\,1\,2\,4\,7\,6\,5\,3} & 1\,0\,2\,4\,7\,6\,5\,3 & 1\,2\,0\,4\,7\,6\,5\,3 & 1\,2\,4\,0\,7\,6\,5\,3 & 1\,2\,4\,7\,0\,6\,5\,3 & 1\,2\,4\,7\,6\,0\,5\,3 & 1\,2\,4\,7\,6\,5\,0\,3 & {1\,2\,4\,7\,6\,5\,3\,0}
\end{tabular}
\end{center}
\end{table*}
\end{example}

\begin{example}\label{SP-labelings-8psk.v2}
If we study the labelings in Example~\ref{SP-labelings-8psk}, we find that the SSP belongs to the first modified Hadamard class ($\matL_\tr{R}=\matN_3$) while the MSP belongs to a different class, i.e.,
\begin{align}
\matL_{\tr{SSP}} =
\matN_3
\begin{bmatrix}
1\,0\,0\\
0\,1\,0\\
1\,0\,1\\
\end{bmatrix},
\quad
\matL_{\tr{MSP}} =
\matL_\tr{R}
\begin{bmatrix}
1\,1\,1\\
0\,1\,0\\
0\,0\,1\\
\end{bmatrix},
\end{align}
where $\matL_\tr{R}\T=[0,1,2,4,7,6,5,3]$ (in integer notation) is the $233$th labeling generated by the MFLSA (see \tabref{Table.MFLSA.m3}). This shows that the NBC does not span all the labelings that follow the SP principle.
\end{example}

\subsection{NBC and BRGC}

Another way of interpreting the result in Theorem~\ref{theo.equivalence} is that for any TCM encoder $\tilde{\Theta}=[\tilde{\matG},\tilde{\matL}]$, a new equivalent TCM encoder can be generated using an encoder $\matG=\tilde{\matG}\matT^{-1}$ and a labeling $\matL=\tilde{\matL}\matT^{-1}$ that belongs to the same modified Hadamard class as the original labeling $\tilde{\matL}$. One direct consequence of this result is that any TCM encoder using the NBC labeling $\matN_m$ and a convolutional encoder $\matG$ is equivalent to a TCM encoder using the BRGC $\matB_m$ and a convolutional encoder $\matG\matT$ with $\matT$ given by \eqref{T_BRGC_NBC}. This is formalized in the following theorem.

\begin{theorem}\label{BRGC_NBC.General}
The BRGC and the NBC of any order $m$ belong to the same modified Hadamard class.
\end{theorem}
\begin{IEEEproof}
The BRGC and NBC are related via $\matB_m = \matN_m \matT$, with $\matT$ given by \eqref{T_BRGC_NBC}. The theorem now follows from Theorem~\ref{LT.Theorem} and the definition of a modified Hadamard class.
\end{IEEEproof}

\begin{example}
\label{NBC_BRGC_m2}
For the two TCM encoders in \figref{TCM_BICM-T_example}, the NBC and BRGC labelings are related via $\matB_2=\matN_2\matT$, i.e.,
\begin{align}
\begin{bmatrix}
0\,0 \\
0\,1 \\
1\,1 \\
1\,0 \\
\end{bmatrix}
=
\begin{bmatrix}
0\,0 \\
0\,1 \\
1\,0 \\
1\,1 \\
\end{bmatrix}
\begin{bmatrix}
1\,1\\
0\,1\\
\end{bmatrix}.
\end{align}
Thus, the BRGC and the NBC of order $m=2$ belong to the same modified Hadamard class, and convolutional encoders can be chosen to make the two resulting TCM encoders equivalent. This was illustrated in \figref{TCM_BICM-T_example}, where the transform block corresponds to the transform matrix $\matT=[1,1;0,1]=\matT^{-1}$. Since $\matN_2=\matB_2\matT^{-1}$, the TCM encoders $[\matG_{[13,17]}, \matB_2]$ and $[\matG_{[13,4]},\matN_2]$ are equivalent, where
\begin{align*}
\matG_{[13,4]}=
\begin{bmatrix}
1\,0\,1\,1\\
0\,1\,0\,0\\
\end{bmatrix}\T
=
\matG_{[13,17]}\matT^{-1}
=
\begin{bmatrix}
1\,0\,1\,1 \\
1\,1\,1\,1\\
\end{bmatrix}\T
\begin{bmatrix}
1\,1\\
0\,1\\
\end{bmatrix}.
\end{align*}
\end{example}


Example~\ref{NBC_BRGC_m2} and Theorem~\ref{BRGC_NBC.General} explain, in part, the results obtained in \cite{Alvarado10d}, where it is shown that the encoders in \cite[Table~III]{Alvarado10d} used with the BRGC perform asymptotically as well as Ungerboeck's TCM.\footnote{The ``in part'' comes from the fact that the system studied in \cite{Alvarado10d} uses a (suboptimal) BICM receiver.}

\section{Error Probability Analysis}\label{Sec:ErroProb}

The results in \secref{Sec:H} are valid for any memoryless channel model and any receiver; however, from now on we focus on the AWGN channel and a maximum likelihood (ML) decoder. In this section, we briefly review bounds on the error probability of TCM encoders under these constraints. These bounds will be used in \secref{Sec:ErroProb.OptimalTCM} to define optimal TCM encoders. The bounds we develop can be found in standard textbooks, see, e.g., \cite[Ch.~4]{Biglieri91_Book} and \cite[Ch.~6]{Schlegel04_Book}, and are re-derived here to make the paper self-contained.

Since TCM encoders are in general not linear\footnote{Note that the usual definition of linearity applies to codes in GF$(q)^N$. However, since TCM codes are defined over the real numbers, the usual definition of linearity does not apply.}, the probability of error depends on the transmitted sequence, i.e., it is not possible to make the assumption that the all-zero sequence was transmitted \cite[p.~101]{Biglieri91_Book}. This constraint can be lifted if the TCM encoder is ``regular'' \cite[Lemma~2]{Calderbank87}, ``superlinear'' \cite[Sec.~II-D]{Benedetto88}, ``scrambled'' \cite{Alvarado10d}, or ``uniform'' \cite{Zehavi87}, \cite[Ch.~18]{Lin04_Book}. However, regularity, superlinearity and uniformity do not hold for all constellation and labelings\footnote{For $8$PSK for example, there is in fact no binary labeling that gives a regular TCM encoder \cite[Sec.~3.3]{Schlegel04_Book}.}, and thus, we cannot use it in this paper.

We consider a baseband-equivalent discrete-time real-valued multi-dimensional AWGN channel. The transmitted sequence of equally likely symbols is denoted by $\un{\bx}=[\bx[1],\ld,\bx[\Ns]]$ where $\bx[n]\in\mcX$ is the $N$-dimensional symbol transmitted at discrete time $n$ and $\Ns$ is the block length. The received sequence of symbols is $\un{\by}=[\by[1],\ld,\by[\Ns]]$, where $\by[n]=\bx[n]+\bz[n]\in\ms{R}^{N}$ is the received vector at time instant $n$. The channel noise $\bz[n]\in\ms{R}^{N}$ is an $N$-dimensional vector with samples of independent and identically distributed (i.i.d.) random variables with zero mean and variance $N_0/2$ per dimension. The signal-to-noise ratio (SNR) is defined as ${\Es}/{N_0}={1}/{N_0}$. The conditional transition PDF of the channel is given by $p_{\bY|\bX}(\by|\bs_q) ={(N_0\pi)^{-\frac{N}{2}}}\exp{-{N_0}^{-1}\|\by-\bs_q\|^2}$.

\subsection{Error Bounds}\label{Sec:ErroProb.Inf}

Let $\mc{X}_{\ell}$ be the set of all length-$\ell$ symbol sequences that start at an arbitrary time instant  and encoder state. Let $\hat{\mc{X}}_{\ell}(\un{\bx})$ be the set of length-$\ell$ sequences $\hat{\un{\bx}}\neq \un{\bx}$ that start and end at the same encoder state as $\un{\bx}\in\mc{X}_{\ell}$ and where all the other $\ell-1$ intermediate states are different. An error event occurs when the decoder chooses a sequence $\hat{\un{\bx}}\in\hat{\mc{X}}_{\ell}(\un{\bx})$ which is different from the transmitted sequence $\un{\bx}$. Using the union bound, the probability of an error event of an ML TCM decoder at a given time instant can be upper-bounded as \cite[eq.~(4.1)]{Biglieri91_Book}\footnote{All the bounds in this section are dependent on the TCM encoder $\Theta$. However, to alleviate the notation, we omit writing out $\Theta$ as an explicit argument.}
\begin{equation}\label{eq:Pse}
P_e\leq \sum_{\ell=1}^\infty\sum_{\un{\bx}\in\mc{X}_{\ell}}P_{\un{\bX}}(\un{\bx})\sum_{\hat{\un{\bx}}\in\hat{\mc{X}}_{\ell}(\un{\bx})}\tr{PEP}(\un{\bx},\hat{\un{\bx}}),
\end{equation}
where $\tr{PEP}(\un{\bx},\hat{\un{\bx}})$ is the pairwise error probability (PEP) and $P_{\un{\bX}}(\un{\bx})$ is the probability that the encoder generates the sequence $\un{\bx}$.

Assuming i.i.d. information bits, the probability of the sequence starting at a given state is $1/2^\nu$. There are $2^{k}$ equally likely branches leaving each state of the trellis at each time instant, and thus,
\begin{equation}\label{P.Seq.x}
P_{\un{\bX}}(\un{\bx})= \frac{1}{2^\nu}\frac{1}{2^{k\ell}}.
\end{equation}
The PEP depends only on the accumulated squared ED (SED) between $\un{\bx}$ and $\hat{\un{\bx}}$ and can be shown to be
\begin{align}
\label{PEP.TCM.AWGN.1}
\tr{PEP}(\un{\bx},\hat{\un{\bx}}) 		& = Q\left(\sqrt{\frac{\Es}{2N_0}\sum_{n=1}^{\ell}\| \bx[n]-\hat{\bx}[n]\|^2}\right).
\end{align}

Let $\Adl$ denote the number of pairs $\un{\bx}\in\mc{X}_{\ell}$ and $\hat{\un{\bx}}\in\hat{\mc{X}}_{\ell}(\un{\bx})$ at accumulated SED $d^2=\sum_{n=1}^{\ell}\| \bx[n]-\hat{\bx}[n]\|^2$ and let $\Awdl$ denote the number of pairs at accumulated SED $d^2$ generated by input sequences at Hamming distance $w$. Using \eqref{P.Seq.x}--\eqref{PEP.TCM.AWGN.1} and the definition of $\Adl$, \eqref{eq:Pse} can be expressed as
\begin{align}
P_e	& \leq 
\label{eq:Pse3}
	\sum_{d^2\in\mc{D}} \Ad Q\left(\sqrt{\frac{d^2\Es}{2N_0}}\right),
\end{align}
 where
\begin{align}\label{Ad}
\Ad \triangleq \sum_{\ell=1}^\infty \frac{1}{2^\nu}\frac{1}{2^{k\ell}} \Adl = \sum_{\ell=1}^\infty \frac{1}{2^\nu}\frac{1}{2^{k\ell}} \sum_{w=1}^{\infty} \Awdl
\end{align}
is the \emph{distance multiplicity} of the TCM encoder. In \eqref{eq:Pse3} $\mc{D}$ is the set of all possible accumulated SEDs between any two sequences, i.e., all the values of $d^2$ for which $\Ad\neq 0$.

To obtain a bound on the BER, each error event must be weighted by the number of bits in error ($w$ out of $k$), \ie
\begin{align}\label{BER}
\tr{BER}&\leq
\sum_{d^2\in\mc{D}} \Bd  Q\left(\sqrt{\frac{d^2\Es}{2N_0}}\right),
\end{align}
where
\begin{align}\label{BdE}
\Bd \triangleq \sum_{\ell=1}^\infty \frac{1}{2^\nu}\frac{1}{2^{k\ell}} \sum_{w=1}^{\infty} \frac{w}{k} \Awdl
\end{align}
is the \emph{bit multiplicity} of the TCM encoder.

Finally, to obtain a bound on the FER we generalize the bound presented in \cite{Caire98b} for convolutional codes to obtain
\begin{align}\label{FER}
\tr{FER}&\leq \Ns \sum_{d^2\in\mc{D}}  \Ad  Q\left(\sqrt{\frac{d^2\Es}{2N_0}}\right).
\end{align}

\subsection{Optimum Distance Spectrum TCM Encoders}\label{Sec:ErroProb.OptimalTCM}

In this section we define TCM encoders that are optimal for asymptotically high SNR. These definitions will be used in \secref{Sec:Results} to tabulate optimized TCM encoders for different configurations.

We call the infinite set of triplets $\{\dE^2,\Ad,\Bd\}$ the distance spectrum (DS) of a given TCM encoder $\Theta=[\matG,\matL]$, where $d^2\in\mc{D}$. We also define the $i$th SED of a given TCM encoder by $d_i^2$ with $i=1,2,3,\ld$, where $d_{i+1}^2>d_{i}^2$ and $d_1^2$ is the minimum SED of the TCM encoder. These SEDs correspond to the ordered set of SEDs in $\mc{D}$. Based on \eqref{BER} and \eqref{FER} we define an optimum DS-TCM (ODS-TCM) as follows.

\begin{definition}
\label{SDS}
A TCM encoder $\Theta=[\matG,\matL]$ with DS $\{\dE^2,\Ad,\Bd\}$ is said to have a superior DS to another TCM encoder $\tilde{\Theta}=[\tilde{\matG},\tilde{\matL}]$ with DS $\{\tilde{d}^2, \tAd,\tBd\}$ if one of the following conditions is fulfilled:
\begin{enumerate}
\item $d_{1}^2 > \tilde{d}_{1}^2$, or
\item $d_{1}^2 = \tilde{d}_{1}^2$, $A_{d_{1}^2}<\tilde{A}_{\tilde{d}_{1}^2}$ and $B_{d_{1}^2}<\tilde{B}_{\tilde{d}_{1}^2}$, or
\item there exist an integer $l> 1$ such that $d_{i}^2=\tilde{d}_{i}^2$, $A_{d_{i}^2}=\tilde{A}_{\tilde{d}_{i}^2}$ and $B_{d_{i}^2}=\tilde{B}_{\tilde{d}_{i}^2}$ for $i=1,2,\ld,l-1$ and $d_{l}^2>\tilde{d}_l^2$ or $d_{l}^2=\tilde{d}_{l}^2$, $A_{d_{l}^2}<\tilde{A}_{\tilde{d}_{l}^2}$ and $B_{d_{l}^2}<\tilde{B}_{\tilde{d}_{l}^2}$.
\end{enumerate}
\end{definition}

\begin{definition}
\label{ODS-TCM}
For a given constellation $\matX$ and memory $\nu$, the TCM encoder $\Theta=[\matG,\matL]$ is said to be an ODS-TCM encoder if no other TCM encoder $\tilde{\Theta}=[\tilde{\matG},\tilde{\matL}]$, for all $\tilde{\matG} \in \mc{G}_{k,m,\nu}$ and $\tilde{\matL} \in \mc{L}_m$, has a superior DS compared to $\Theta$.
\end{definition}

An ODS-TCM encoder in Definition~\ref{ODS-TCM} is the asymptotically optimal TCM encoder in terms of BER and FER for a given block length $\Ns$. Unlike the more classical definition of optimal encoders, ODS-TCM encoders are defined as encoders that are optimal in terms of \emph{both} $\Ad$ and $\Bd$. This implies that in principle, for some combinations of $k,m,\nu$, it is possible that no ODS-TCM encoder exists. As we will see in \secref{Sec:Results}, this is not an uncommon situation. Moreover, by using this somehow nonstandard definition we avoid listing encoders that have optimal BER performance but possibly rather poor FER performance (or vice versa). This situation happens for $R=1/2$ and $4$PAM, as we will show in \secref{Sec:Results.PAM}.

\section{Numerical Results}\label{Sec:Results}

In this section we study well-structured one- and two-dimensional constellations, i.e., $M$PAM and $M$PSK constellations. An $M$PAM constellation is defined by $\matX =[x_1,x_2,\ld,x_M]\T$ with $x_q=-(M+1-2q)\Delta\in \ms{R}$, $q=1,\ld,M$, and $\Delta^2=3/(M^2-1)$ so that $\Es =1$. An $M$PSK constellation is defined by $\matX=[\bx_1;\bx_2;\ld;\bx_M]$ with $\bx_q=[\cos{(2\pi (q-1)/M)},\sin{(2\pi (q-1)/M)}]\in \ms{R}^2$ and $q=1,\ld,M$.

In the following sections we show results of exhaustive searches over $\mc{G}_{k,m,\nu}$ and $\mc{R}_m$, and thus, these results should be understood as a complete answer to the problem of jointly designing the feedforward encoder and the labeling for TCM encoders. The ODS-TCM encoders presented are obtained by comparing the first five nonzero elements in the spectrum, which we numerically calculate using a generalization of the algorithm presented in \cite[Sec.~12.4.3]{Benedetto99_Book}.\footnote{Note that if more than five elements are considered different ODS-TCM encoders might be found.} On the other hand, the bounds used to compare with simulation results were calculated using 20 terms. The tabulated results are ordered first in terms of the output of the MFLSA, then in lexicographic order for the memories $\nu_1,\ld,\nu_k$, and then in lexicographic order for the encoder matrices $\matG$. This ordering becomes relevant when there are multiple TCM encoder with identical (and optimal) five-term DS.

\subsection{ODS-TCM Encoders for $M$PAM}\label{Sec:Results.PAM}

$M$PAM constellations are symmetric around zero. Because of this, two TCM encoders based on an $M$PAM constellation, the first one using the labeling $\matL=[\bc_1;\bc_2;\ld;\bc_{M-1};\bc_M]$ and the second one using a ``reverse'' labeling $\matL'=[\bc_M;\bc_{M-1};\ld;\bc_2;\bc_1]$, are equivalent for any $M$. This result implies that the number of binary labelings that give nonequivalent TCM encoders is $M_\tr{R}/2$.
Specifically, for $m=2$ and $m=3$ ($4$PAM and $8$PAM), only $2$ and $120$ labelings need to be evaluated, respectively, instead of 24 and 40320 in an exhaustive search, see \tabref{Table_Mi_Me}.

To generate only the $M_\tr{R}/2$ nonequivalent labelings for $M$PAM, the MFLSA in Algorithm~1 can be modified as follows. Replace $M$ on lines 5 and 6 with $e(\ind)$, where the integer function $e(q)$ is defined as $M/2$ if $q=0$ and $M$ otherwise. This has the effect of only generating labelings in which the all-zero label is among the first $M/2$ positions (i.e., the first 4 columns of \tabref{Table.MFLSA.m3} for $8$PAM).

\subsubsection{$R=1/2$ and $4$PAM}\label{Sec:Results.PAM.4}
\begin{figure}[!tb]
\newcommand{\scale}{0.82}
\psfrag{xlabel}[cc][cc][\scale]{$\Es/N_0$~[dB]}
\psfrag{ylabel}[bc][Bc][\scale]{FER/BER}
\psfrag{Bound FER}[cl][cl][\scale]{FER Bound}
\psfrag{Bound BER}[cl][cl][\scale]{BER Bound}
\psfrag{Sim. UNG}[cl][cl][\scale]{Sim. $[\cd]^\tr{U}$}
\psfrag{Sim. NEW}[cl][cl][\scale]{Sim. $[\cd]^\tr{AB}$}
\psfrag{K=5}[cc][cc][\scale]{$\nu=4$}
\psfrag{K=7}[cr][cr][\scale]{$\nu=6$}
\begin{center}
	\includegraphics{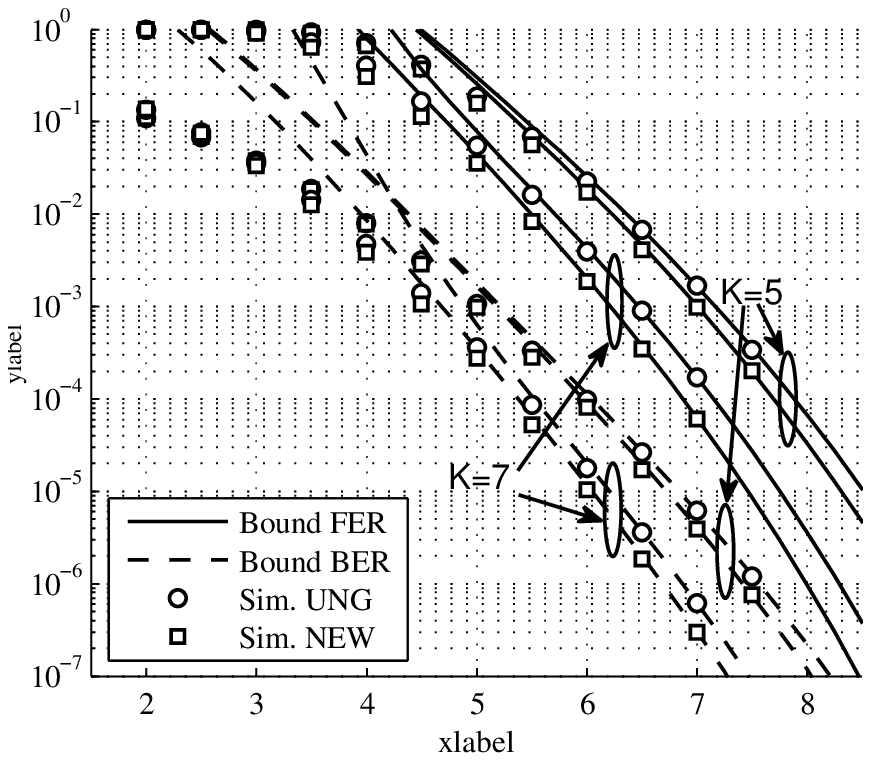}
	\caption{BER/FER bounds in \eqref{BER} and \eqref{FER} and simulations for Ungerboeck's encoders and the ODS-TCM encoders in \tabref{tab:R12-4PAM} for $\Ns=1000$, $4$PAM, $R=1/2$ (1~[bit/symbol]), and $\nu=4,6$.}
    \label{tcm_k_1_n_2_Ns_1000}
\end{center}
\end{figure}

The results obtained for $R=1/2$ and $4$PAM and different values of $\nu$ are shown in \tabref{tab:R12-4PAM}. The table reports the DS as well as the labeling and convolutional encoder for the ODS-TCM encoders (shown as $[\cd]^{\tr{AB}}$). For $\nu=5$, however, no ODS-TCM encoder was found, i.e., there is no TCM encoder that is optimal in terms of both $\Ad$ and $\Bd$. Instead, we list the TCM encoder with best $\Ad$ among those with optimal $\Bd$ (shown as $[\cd]^{\tr{B}}$), or vice versa (shown as $[\cd]^{\tr{A}}$). In this table we also include Ungerboeck's encoders\footnote{Ungerboeck did not report results for $\nu=1$, and thus, we do not include them in the Tables, \ie we only show the ODS-TCM encoder for $\nu=1$.}, which we denote by $[\cd]^\tr{U}$. When Ungerboeck's labeling (NBC) or Ungerboeck's convolutional encoder coincide with $[\cd]^\tr{AB}$ or $[\cd]^\tr{B}$, we use the notation $[\cd]^\tr{UAB}$ or $[\cd]^\tr{UB}$, respectively. The results in \tabref{tab:R12-4PAM} show that no gains in terms of MED are obtained and that the NBC is indeed the optimal labeling for all memories. The key difference between Ungerboeck's design and the ODS-TCM encoders is the better multiplicities obtained. To compare the gains obtained by the ODS-TCM encoders over Ungerboeck's encoders, we show in \figref{tcm_k_1_n_2_Ns_1000} their BER/FER for $\nu=4,6$. This figure clearly shows the gains obtained by using the ODS-TCM encoders which are visible not only at high SNR, but also for low SNR values (see, \eg the FER markers for $\nu=6$).

\begin{table*}[thb]
\scriptsize\centering
\renewcommand{\arraystretch}{1.4}
\caption{Distance spectrum of ODS-TCM encoders ($[\cd]^\tr{AB}$) and Ungerboeck's encoders ($[\cdot]^\textrm{U}$) for $k=1$~[bit/symbol] and $4$PAM ($m=2$). The notation $[\cd]^\tr{A}$ and $[\cd]^\tr{B}$ is used when no ODS-TCM encoder was found.}
\label{tab:R12-4PAM}
\begin{tabular}{@{~}c@{~}|@{~}l@{~}|@{~}l@{~}|@{~}l@{~}l@{~}l@{~}l@{~}l@{~}}
\hline
\hline
$\nu$ & \hspace{6mm}$\matL\T$ & \hspace{6mm}$\matG$ & \multicolumn{5}{@{~}l@{~}}{Distance Spectrum $\{\dE^2,\Ad,\Bd\}$}\\
\hline
\hline
$1$ & $[0,\!1,\!2,\!3]^\textrm{AB~}\!$ & $[3,\!1]^\textrm{AB~}\!$ & \{4.00,\,0.50,\,0.50\}, & \{4.80,\,0.50,\,1.00\}, & \{5.60,\,0.50,\,1.50\}, & \{6.40,\,0.50,\,2.00\}, & \{7.20,\,0.50,\,2.50\} \\
\hline
\multirow{2}{*}{$2$} & \multirow{2}{*}{$[0,\!1,\!2,\!3]^\textrm{UAB}\!$} & $[5,\!2]^\textrm{U~~}\!$ & \{7.20,\,1.00,\,1.00\}, & \{8.00,\,1.25,\,2.50\}, & \{8.80,\,1.75,\,5.25\}, & \{9.60,\,2.56,\,10.25\}, & \{10.40,\,3.81,\,19.06\} \\
                   &  & $[7,\!2]^\textrm{AB~}\!$ & \{7.20,\,0.50,\,0.50\}, & \{8.00,\,1.25,\,2.50\}, & \{8.80,\,1.63,\,4.88\}, & \{9.60,\,2.56,\,10.25\}, & \{10.40,\,3.78,\,18.91\} \\
\hline
$3$ & $[0,\!1,\!2,\!3]^\textrm{UAB}\!$ & $[13,\!4]^\textrm{UAB}\!$ & \{8.00,\,0.25,\,0.50\}, & \{8.80,\,1.00,\,3.00\}, & \{9.60,\,1.56,\,6.25\}, & \{10.40,\,2.75,\,9.75\}, & \{11.20,\,3.14,\,16.84\} \\
\hline
\multirow{2}{*}{$4$} & \multirow{2}{*}{$[0,\!1,\!2,\!3]^\textrm{UAB}\!$} & $[23,\!4]^\textrm{U~~}\!$ & \{8.80,\,0.63,\,1.88\}, & \{9.60,\,0.50,\,2.00\}, & \{10.40,\,2.00,\,6.00\}, & \{11.20,\,2.02,\,10.09\}, & \{12.00,\,2.03,\,13.22\} \\
                   &  & $[23,\!10]^\textrm{AB~}\!$ & \{8.80,\,0.13,\,0.38\}, & \{9.60,\,0.50,\,2.00\}, & \{10.40,\,1.88,\,5.38\}, & \{11.20,\,2.39,\,10.34\}, & \{12.00,\,3.72,\,21.03\} \\
\hline
\multirow{2}{*}{$5$} & \multirow{2}{*}{$[0,\!1,\!2,\!3]^\textrm{UAB}\!$} & $[45,\!10]^\textrm{UB~}\!$ & \{10.40,\,1.13,\,1.63\}, & \{11.20,\,1.52,\,5.09\}, & \{12.00,\,2.59,\,12.16\}, & \{12.80,\,3.58,\,22.13\}, & \{13.60,\,5.29,\,38.60\} \\
                   &  & $[55,\!4]^\textrm{A~~}\!$ & \{10.40,\,0.75,\,1.75\}, & \{11.20,\,2.13,\,8.75\}, & \{12.00,\,2.14,\,10.48\}, & \{12.80,\,4.47,\,24.75\}, & \{13.60,\,5.45,\,37.01\} \\
\hline
\multirow{2}{*}{$6$} & \multirow{2}{*}{$[0,\!1,\!2,\!3]^\textrm{UAB}\!$} & $[103,\!24]^\textrm{U~~}\!$ & \{11.20,\,2.34,\,5.91\}, & \{12.80,\,2.82,\,22.01\}, & \{14.40,\,7.60,\,57.35\}, & \{16.00,\,31.39,\,268.35\}, & \{17.60,\,74.37,\,779.76\} \\
                   &  & $[107,\!32]^\textrm{AB~}\!$ & \{11.20,\,0.13,\,0.50\}, & \{12.00,\,1.44,\,5.81\}, & \{12.80,\,1.41,\,5.77\}, & \{13.60,\,1.73,\,12.58\}, & \{14.40,\,4.58,\,31.53\} \\
\hline
\multirow{2}{*}{$7$} & \multirow{2}{*}{$[0,\!1,\!2,\!3]^\textrm{UAB}\!$} & $[235,\!126]^\textrm{U~~}\!$ & \{12.80,\,2.19,\,8.19\}, & \{14.40,\,3.05,\,17.66\}, & \{16.00,\,10.09,\,89.43\}, & \{17.60,\,25.03,\,231.04\}, & \{19.20,\,90.45,\,920.63\} \\
                   &  & $[313,\!126]^\textrm{AB~}\!$ & \{12.80,\,1.46,\,8.02\}, & \{14.40,\,4.77,\,34.60\}, & \{16.00,\,15.42,\,130.51\}, & \{17.60,\,35.60,\,375.08\}, & \{19.20,\,103.30,\,1213.89\} \\
\hline
\multirow{2}{*}{$8$} & \multirow{2}{*}{$[0,\!1,\!2,\!3]^\textrm{UAB}\!$} & $[515,\!362]^\textrm{U~~}\!$ & \{13.60,\,0.53,\,4.66\}, & \{14.40,\,1.89,\,10.79\}, & \{15.20,\,1.66,\,14.10\}, & \{16.00,\,3.81,\,30.45\}, & \{16.80,\,6.03,\,49.34\} \\
                   &  & $[677,\!362]^\textrm{AB~}\!$ & \{13.60,\,0.36,\,2.05\}, & \{14.40,\,1.06,\,6.41\}, & \{15.20,\,1.47,\,11.09\}, & \{16.00,\,3.44,\,23.69\}, & \{16.80,\,5.25,\,41.32\} \\
\hline
\hline
\end{tabular}
\end{table*}

\subsubsection{$R=2/3$ and $8$PAM}\label{Sec:Results.PAM.8}
\begin{figure}[!tb]
\newcommand{\scale}{0.82}
\psfrag{xlabel}[cc][cc][\scale]{$\Es/N_0$~[dB]}
\psfrag{ylabel}[bc][Bc][\scale]{FER/BER}
\psfrag{Bound FER}[cl][cl][\scale]{FER Bound}
\psfrag{Bound BER}[cl][cl][\scale]{BER Bound}
\psfrag{Sim. UNG}[cl][cl][\scale]{Sim. $[\cd]^\tr{U}$}
\psfrag{Sim. NEW}[cl][cl][\scale]{Sim. $[\cd]^\tr{AB}$}
\psfrag{K=5}[cr][cr][\scale]{$\nu=4$}
\psfrag{K=7}[cr][cr][\scale]{$\nu=6$}
\begin{center}
	\includegraphics{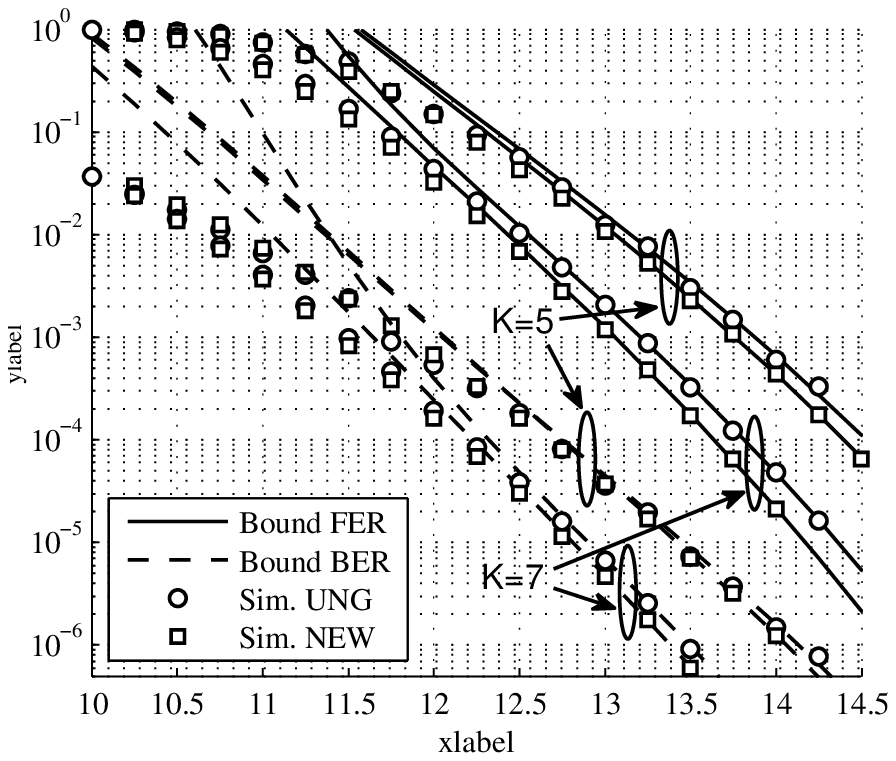}
	\caption{BER/FER bounds in \eqref{BER} and \eqref{FER} and simulations for Ungerboeck's encoders and the ODS-TCM encoders in \tabref{tab:R23-8PAM} for $\Ns=1000$, $8$PAM, $R=2/3$ (2~[bit/symbol]), and $\nu=4,6$.}
    \label{tcm_k_2_n_3_Ns_1000}
\end{center}
\end{figure}

The results for $R=2/3$ and $8$PAM are shown in \tabref{tab:R23-8PAM}. For $\nu=1,2,3,4,6$ the reported encoders are in the form $[\cd]^{\tr{AB}}$, while for $\nu=5$ no ODS-TCM was found, and we use the same notation as for $4$PAM. Unlike for $R=1/2$, the parity-check matrix reported by Ungerboeck for $R=2/3$ specifies the code but not the encoder. To have a fair comparison between Ungerboeck's codes with the ODS-TCM encoders, we first listed all the convolutional encoders that give Ungerboeck's parity-check matrix and then pick the one with optimal $\Bd$ (all of them have the same $\Ad$). These are the encoders reported in \tabref{tab:R23-8PAM} as $[\cd]^{\tr{U}}$. Even though Ungerboeck's encoders in \tabref{tab:R23-8PAM} are the best encoders for that particular parity-check matrix, they coincide with the $[\cd]^{\tr{B}}$ encoders only for one out of six cases ($\nu=5$). For all the other cases, the ODS-TCM encoders result in a better spectrum. Also, unlike for $4$PAM, \tabref{tab:R23-8PAM} shows that the NBC is not the optimal labeling. For example, for $\nu=4$, the optimal labeling is $\matL\T=[1, 2, 4, 0, 6, 5, 3, 7]^\tr{AB}$, which does not follow the SP principle (\cf Definition~\ref{SP.definition}). In \figref{tcm_k_2_n_3_Ns_1000}, we show the BER/FER results obtained by the ODS-TCM encoders for $R=2/3$, $8$PAM, and $\nu=4,6$. This figure shows the tightness of the bounds and again gains over Ungerboeck's encoders.

\begin{table*}[thb]
\scriptsize\centering
\renewcommand{\arraystretch}{1.4}
\caption{Distance spectrum of ODS-TCM encoders ($[\cd]^\tr{AB}$) and Ungerboeck's encoders ($[\cdot]^\textrm{U}$) for $k=2$~[bit/symbol] and $8$PAM ($m=3$). The notation $[\cd]^\tr{A}$ and $[\cd]^\tr{B}$ is used when no ODS-TCM encoder was found.}
\label{tab:R23-8PAM}
\begin{tabular}{@{~}c@{~}|@{~}l@{~}|@{~}l@{~}|@{~}l@{~}l@{~}l@{~}l@{~}l@{~}}
\hline
\hline
$\nu$ & \hspace{10mm}$\matL\T$ & \hspace{10mm}$\matG$ & \multicolumn{5}{@{~}l@{~}}{Distance Spectrum $\{\dE^2,\Ad,\Bd\}$}\\
\hline
\hline
$1$ & $[1,\!2,\!4,\!0,\!6,\!5,\!3,\!7]^\textrm{AB~}\!$ & $[1,\!1,\!1;1,\!3,\!0]^\textrm{AB~}\!$ & \{0.95,\,1.13,\,0.84\}, & \{1.14,\,1.13,\,1.69\}, & \{1.33,\,1.13,\,2.53\}, & \{1.52,\,1.13,\,3.38\}, & \{1.71,\,1.13,\,4.22\} \\
\hline
\multirow{2}{*}{$2$} & \multirow{2}{*}{$[0,\!1,\!2,\!3,\!4,\!5,\!6,\!7]^\textrm{UAB}\!$} & $[1,\!0,\!0;0,\!5,\!2]^\textrm{U~~}\!$ & \{1.71,\,2.25,\,1.88\}, & \{1.90,\,3.52,\,5.11\}, & \{2.10,\,6.05,\,12.35\}, & \{2.29,\,10.56,\,27.64\}, & \{2.48,\,18.47,\,58.91\} \\
                   &  & $[1,\!0,\!0;0,\!7,\!2]^\textrm{AB~}\!$ & \{1.71,\,1.69,\,1.69\}, & \{1.90,\,3.52,\,5.11\}, & \{2.10,\,6.01,\,12.34\}, & \{2.29,\,10.56,\,27.64\}, & \{2.48,\,18.46,\,58.91\} \\
\hline
\multirow{2}{*}{$3$} & $[0,\!1,\!2,\!3,\!4,\!5,\!6,\!7]^\textrm{U~~}\!$ & $[1,\!0,\!0;0,\!13,\!4]^\textrm{U~~}\!$ & \{1.90,\,1.27,\,2.11\}, & \{2.10,\,3.38,\,6.75\}, & \{2.29,\,5.49,\,14.14\}, & \{2.48,\,12.45,\,32.48\}, & \{2.67,\,18.59,\,64.81\} \\
                   & $[1,\!2,\!4,\!0,\!6,\!5,\!3,\!7]^\textrm{AB~}\!$ & $[1,\!1,\!1;2,\!15,\!0]^\textrm{AB~}\!$ & \{1.90,\,1.27,\,1.90\}, & \{2.10,\,3.38,\,8.44\}, & \{2.29,\,5.49,\,17.25\}, & \{2.48,\,12.45,\,38.50\}, & \{2.67,\,18.59,\,74.81\} \\
\hline
\multirow{2}{*}{$4$} & $[0,\!1,\!2,\!3,\!4,\!5,\!6,\!7]^\textrm{U~~}\!$ & $[1,\!0,\!0;0,\!23,\!4]^\textrm{U~~}\!$ & \{2.10,\,2.64,\,5.59\}, & \{2.29,\,2.53,\,6.75\}, & \{2.48,\,6.75,\,13.50\}, & \{2.67,\,12.11,\,40.55\}, & \{2.86,\,15.99,\,66.51\} \\
                   & $[1,\!2,\!4,\!0,\!6,\!5,\!3,\!7]^\textrm{AB~}\!$ & $[1,\!1,\!1;2,\!31,\!0]^\textrm{AB~}\!$ & \{2.10,\,0.95,\,1.90\}, & \{2.29,\,2.53,\,7.59\}, & \{2.48,\,7.91,\,21.78\}, & \{2.67,\,13.21,\,45.70\}, & \{2.86,\,19.77,\,88.01\} \\
\hline
\multirow{2}{*}{$5$} & \multirow{2}{*}{$[0,\!1,\!2,\!3,\!4,\!5,\!6,\!7]^\textrm{UAB}\!$} & $[1,\!0,\!0;0,\!45,\!10]^\textrm{UB~}\!$ & \{2.48,\,4.32,\,6.54\}, & \{2.67,\,7.99,\,19.45\}, & \{2.86,\,14.26,\,46.29\}, & \{3.05,\,27.05,\,102.83\}, & \{3.24,\,44.27,\,201.33\} \\
                   &  & $[1,\!0,\!0;0,\!55,\!4]^\textrm{A~~}\!$ & \{2.48,\,3.80,\,6.96\}, & \{2.67,\,8.74,\,21.63\}, & \{2.86,\,13.53,\,45.10\}, & \{3.05,\,29.51,\,106.50\}, & \{3.24,\,44.49,\,198.08\} \\
\hline
\multirow{2}{*}{$6$} & \multirow{2}{*}{$[0,\!1,\!2,\!3,\!4,\!5,\!6,\!7]^\textrm{UAB}\!$} & $[1,\!0,\!0;0,\!103,\!24]^\textrm{U~~}\!$ & \{2.67,\,10.74,\,22.97\}, & \{3.05,\,19.91,\,86.93\}, & \{3.43,\,72.68,\,343.40\}, & \{3.81,\,353.99,\,1927.40\}, & \{4.19,\,1137.86,\,7442.94\} \\
                   &  & $[1,\!0,\!0;0,\!107,\!32]^\textrm{AB~}\!$ & \{2.67,\,1.42,\,4.27\}, & \{2.86,\,8.46,\,24.43\}, & \{3.05,\,12.94,\,40.47\}, & \{3.24,\,15.68,\,74.20\}, & \{3.43,\,40.61,\,182.47\} \\
\hline
\hline
\end{tabular}
\end{table*}

\subsection{ODS-TCM Encoders for $M$PSK}\label{Sec:Results.PSK}

A TCM encoder based on an $M$PSK constellation is not affected by a circular rotation of its labeling, i.e., without loss of generality it can be assumed that the all zero label is assigned to the constellation point $\bx_1=[1,0]$. The consequence of this is that for $M$PSK constellations, the number of reduced column echelon matrices that give nonequivalent TCM encoders is further reduced by a factor of $M$. In view of the results in \tabref{Table_Mi_Me}, for $4$PSK, there is only one labeling that needs to be tested, \eg the NBC. For $m\geq3$, the nonequivalent labelings can be obtained from the MFLSA by setting $\ind \leftarrow 3$ in line 4, which gives the FLSA of \cite{Knagenhjelm96}. For example, for $M=8$, the output corresponds to the first column of \tabref{Table.MFLSA.m3}, which gives 30 labelings.

\subsubsection{$R=1/2$ and $4$PSK}
In this case there is only one labeling to be tested (the NBC), and thus, only a search over the encoders needs to be performed. Moreover, without loss of generality, we can use the BRGC instead (because it is in the same Hadamard class as the NBC) and search over encoders for this labeling. Since $4$PSK with the BRGC can be considered as two independent $2$PAM constellations (one in each dimension), the design of TCM encoders in this case boils down to selecting convolutional encoders with optimal spectrum (in the sense of Definition~\ref{ODS-TCM}).

We have performed an exhaustive search for convolutional encoders with optimal spectrum up to $\nu=12$ and found that our results coincide with those reported in \cite[Table~I]{Sone01}. For $\nu=1,2,3,4,5,6,11,12$ the optimal convolutional encoders ($[\cd]^\tr{AB}$) are in fact the encoders from \cite[Table~I]{Frenger99} (which were initially optimized only in terms of $\Bd$). For $\nu=7,8,9,10$ we found that no optimal encoder exists, i.e., the convolutional encoders optimal in terms of $\Ad$ are not optimal in terms of $\Bd$ and vice versa.\footnote{Convolutional encoders with optimal $\Ad$ and memories up to $\nu=26$ have been recently published in \cite[Table 7.1]{Hug12_Thesis}.} These encoders are in fact shown in \cite[Table~I]{Sone01}\footnote{Although the search in \cite{Sone01} was performed only considering events at minimum Hamming distance and not over the whole spectrum.}, which extends the results in \cite{Frenger99,Chang97,Bocharova97} because it considers both $\Ad$ and $\Bd$ as optimization criteria.

Based on the discussion above, we conclude that an ODS-TCM encoders  can be constructed by concatenating the encoders in \cite[Table~I]{Sone01} with a $4$PSK constellation labeled by the BRGC. Alternatively, ODS-TCM encoders can be obtained by using a $4$PSK constellation labeled by the NBC and using the encoders in \cite[Table~I]{Sone01} after applying the transformation $\matT^{-1}=[1,1;0,1]$. For example, for $\nu=8$, we found $\matG_{[515,677]}$ and $\matG_{[435,657]}$ to be the optimal encoders in terms of $\Ad$ and $\Bd$, respectively, and thus, the two pairs of equivalent ODS-TCM encoders are $\Theta=[\matG_{[515,677]},\matB_2]$ and $\tilde{\Theta}=[\matG_{[515,677]}\matT^{-1},\matN_2]$, and $\Theta=[\matG_{[435,657]},\matB_2]$ and $\tilde{\Theta}=[\matG_{[435,657]}\matT^{-1},\matN_2]$.


\subsubsection{$R=2/3$ and $8$PSK}
The results obtained for $R=2/3$ and $8$PSK are shown in \tabref{tab:R23-8PSK}. Somehow disappointingly, this table shows that the NBC is indeed the optimal labeling in all the cases, and thus, the selection of the labeling for this particular configuration does not provide any gains over Ungerboeck's TCM schemes. The better spectrum obtained by the ODS-TCM encoders in this case then comes only from the selection of the convolutional encoder.

\begin{table*}[thb]
\scriptsize\centering
\renewcommand{\arraystretch}{1.4}
\caption{Distance spectrum of ODS-TCM encoders ($[\cd]^\tr{AB}$) and Ungerboeck's encoders ($[\cdot]^\textrm{U}$) for $k=2$~[bit/symbol] and $8$PSK ($m=3$). The notation $[\cd]^\tr{A}$ and $[\cd]^\tr{B}$ is used when no ODS-TCM encoder was found.}
\label{tab:R23-8PSK}
\begin{tabular}{@{~}c@{~}|@{~}l@{~}|@{~}l@{~}|@{~}l@{~}l@{~}l@{~}l@{~}l@{~}}
\hline
\hline
$\nu$ & \hspace{10mm}$\matL\T$ & \hspace{10mm}$\matG$ & \multicolumn{5}{@{~}l@{~}}{Distance Spectrum $\{\dE^2,\Ad,\Bd\}$}\\
\hline
\hline
$1$ & $[0,\!1,\!2,\!3,\!4,\!5,\!6,\!7]^\textrm{AB~}\!$ & $[1,\!0,\!0;0,\!1,\!2]^\textrm{AB~}\!$ & \{2.59,\,2.00,\,1.50\}, & \{3.17,\,2.00,\,3.00\}, & \{3.76,\,2.00,\,4.50\}, & \{4.00,\,1.00,\,0.50\}, & \{4.34,\,2.00,\,6.00\} \\
\hline
$2$ & $[0,\!1,\!2,\!3,\!4,\!5,\!6,\!7]^\textrm{UAB}\!$ & $[1,\!0,\!0;0,\!5,\!2]^\textrm{UAB}\!$ & \{4.00,\,1.00,\,0.50\}, & \{4.59,\,4.00,\,4.00\}, & \{5.17,\,8.00,\,14.00\}, & \{5.76,\,16.00,\,38.00\}, & \{6.34,\,32.00,\,96.00\} \\
\hline
\multirow{2}{*}{$3$} & \multirow{2}{*}{$[0,\!1,\!2,\!3,\!4,\!5,\!6,\!7]^\textrm{UAB}\!$} & $[1,\!2,\!0;4,\!1,\!2]^\textrm{U~~}\!$ & \{4.59,\,2.00,\,2.50\}, & \{5.17,\,4.00,\,8.50\}, & \{5.76,\,8.00,\,25.00\}, & \{6.00,\,1.00,\,0.50\}, & \{6.34,\,16.00,\,66.00\} \\
                   &  & $[1,\!2,\!0;4,\!5,\!2]^\textrm{AB~}\!$ & \{4.59,\,2.00,\,2.00\}, & \{5.17,\,4.00,\,8.50\}, & \{5.76,\,8.00,\,25.00\}, & \{6.00,\,1.00,\,0.50\}, & \{6.34,\,16.00,\,66.00\} \\
\hline
\multirow{3}{*}{$4$} & \multirow{3}{*}{$[0,\!1,\!2,\!3,\!4,\!5,\!6,\!7]^\textrm{UAB}\!$} & $[2,\!7,\!0;7,\!3,\!2]^\textrm{U~~}\!$ & \{5.17,\,2.25,\,5.50\}, & \{5.76,\,4.63,\,14.13\}, & \{6.00,\,1.00,\,0.50\}, & \{6.34,\,6.06,\,26.50\}, & \{6.59,\,4.00,\,5.50\} \\
                   &  & $[2,\!7,\!0;7,\!1,\!2]^\textrm{A~~}\!$ & \{5.17,\,2.25,\,5.00\}, & \{5.76,\,3.88,\,11.56\}, & \{6.00,\,1.00,\,0.50\}, & \{6.34,\,9.56,\,38.81\}, & \{6.59,\,4.00,\,5.50\} \\
                   &  & $[1,\!4,\!2;6,\!1,\!0]^\textrm{B~~}\!$ & \{5.17,\,2.50,\,5.00\}, & \{5.76,\,3.75,\,11.25\}, & \{6.34,\,8.13,\,32.44\}, & \{6.59,\,3.50,\,4.50\}, & \{6.93,\,16.19,\,80.94\} \\
\hline
\multirow{2}{*}{$5$} & \multirow{2}{*}{$[0,\!1,\!2,\!3,\!4,\!5,\!6,\!7]^\textrm{UAB}\!$} & $[1,\!2,\!0;30,\!25,\!16]^\textrm{U~~}\!$ & \{5.76,\,4.00,\,10.50\}, & \{6.00,\,1.00,\,0.50\}, & \{6.34,\,4.00,\,16.25\}, & \{6.93,\,4.00,\,24.13\}, & \{7.17,\,3.00,\,7.50\} \\
                   &  & $[1,\!2,\!0;30,\!25,\!10]^\textrm{AB~}\!$ & \{5.76,\,2.00,\,5.75\}, & \{6.00,\,1.00,\,0.50\}, & \{6.34,\,3.63,\,15.56\}, & \{6.59,\,3.00,\,5.50\}, & \{6.93,\,8.06,\,40.63\} \\
\hline
\multirow{3}{*}{$6$} & \multirow{3}{*}{$[0,\!1,\!2,\!3,\!4,\!5,\!6,\!7]^\textrm{UAB}\!$} & $[4,\!11,\!0;13,\!4,\!6]^\textrm{U~~}\!$ & \{6.34,\,5.25,\,22.56\}, & \{7.17,\,10.00,\,28.88\}, & \{7.51,\,14.53,\,98.50\}, & \{8.00,\,3.00,\,3.75\}, & \{8.34,\,38.56,\,199.78\} \\
                   &  & $[1,\!6,\!0;27,\!25,\!12]^\textrm{A~~}\!$ & \{6.34,\,3.25,\,12.00\}, & \{7.17,\,7.25,\,17.88\}, & \{7.51,\,19.13,\,119.17\}, & \{8.00,\,3.00,\,5.00\}, & \{8.34,\,36.69,\,159.69\} \\
                   &  & $[1,\!6,\!0;35,\!31,\!6]^\textrm{B~~}\!$ & \{6.34,\,3.56,\,11.50\}, & \{7.17,\,7.25,\,16.88\}, & \{7.51,\,16.58,\,92.05\}, & \{8.00,\,3.50,\,4.75\}, & \{8.34,\,30.63,\,150.81\} \\
\hline
\hline
\end{tabular}
\end{table*}

\begin{figure}[!tb]
\newcommand{\scale}{0.82}
\psfrag{xlabel}[cc][cc][\scale]{$d_i^2$}
\psfrag{ylabel}[Bc][Bc][\scale]{Distance Spectrum}
\psfrag{Ad}[cl][cl][\scale]{$\Ad$}
\psfrag{Bd}[cl][cl][\scale]{$\Bd$}
\psfrag{OAd}[cl][cl][\scale]{$[\cd]^\tr{A}$}
\psfrag{OBd}[cl][cl][\scale]{$[\cd]^\tr{B}$}
\psfrag{OGU}[cl][cl][\scale]{$[\cd]^\tr{U}$}
\begin{center}
	\includegraphics{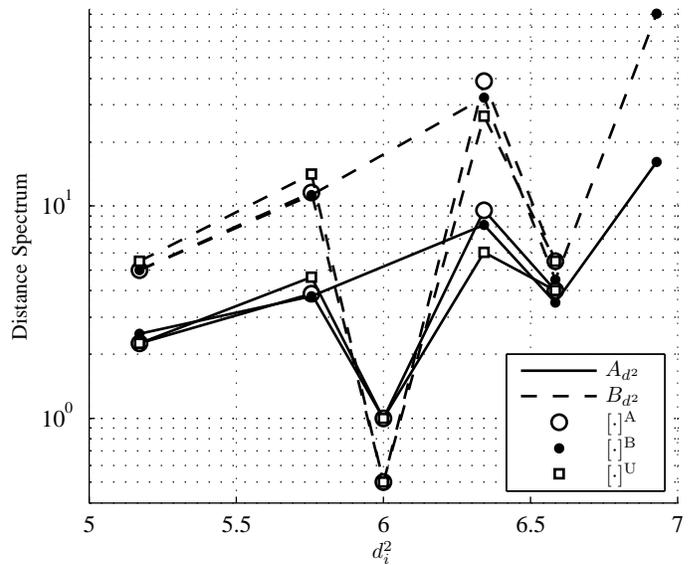}
	\caption{DS for encoders with $\nu=4$ for $R=2/3$ and $8$PSK from \tabref{tab:R23-8PSK}.}
    	\label{spectra_2_3_8PSK_nu_4}
\end{center}
\end{figure}

In \figref{spectra_2_3_8PSK_nu_4}, we show the DS for the encoders in \tabref{tab:R23-8PSK} with $\nu=4$. It is clear from the figure that an encoder optimal in terms of $\Ad$ can be suboptimal in terms of $\Bd$, and vice versa. In addition, the figure shows how the set of SEDs $\mc{D}$ is in general different for different encoders. It also shows how Ungerboeck's encoder is optimal in terms of $\Ad$ for the term at MED, but in general suboptimal if the whole DS is considered. 

We note that depending on $\nu$, the ODS-TCM encoders in \tabref{tab:R23-8PSK} have inferior, equivalent, or superior $\Bd$ spectrum to those listed in \cite[Table~3.2]{Schlegel04_Book}, \cite[Table~6.10]{Zhang96_Thesis}.\footnote{To have a fair comparison, the values of $\Bd$ listed in \cite[Table~3.2]{Schlegel04_Book}, \cite[Table~6.10]{Zhang96_Thesis} should be scaled by a factor $1/k=1/2$.} The reason for this is that the codes tabulated in \cite[Table~3.2]{Schlegel04_Book}, \cite[Table~6.10]{Zhang96_Thesis} are found by searching over parity check matrices and then converted to feedback encoders (in observer canonical form \cite[Fig.~2.2]{Zhang96_Thesis}). On the other hand, we search over a different set of encoders, namely, over all the noncatastrophic feedforward encoders.

All labelings we found for the ODS-TCM encoders (\ie the highlighted labelings in \tabref{Table.MFLSA.m3} and the optimal ones in Tables~\ref{tab:R23-8PAM} and \ref{tab:R23-8PSK}) have optimal EP. This makes us conjecture that good TCM encoders can be found by using the EP of \cite{Wesel01} on top of the proposed classification. This approach would indeed reduce the search space (for example, for $8$PAM and $8$PSK constellations, only eight and two labelings, respectively, would need to be tested). However, it would not allow us to claim optimality in the sense of Definition~\ref{ODS-TCM}.

\section{Conclusions}\label{Sec:Conclusion}

In this paper we analyzed the problem of jointly designing the feedforward convolutional encoder and the labeling of a TCM encoder. It was shown that the number of labelings that need to be checked can be reduced if they are grouped into modified Hadamard classes. This classification allowed us to prove that it is always possible to design a TCM encoder based on the BRGC with identical performance to the one proposed by Ungerboeck in 1982. The numerical results show that in most cases, the NBC is the optimal binary labeling for TCM encoders and that gains up to 0.3~dB over the previously best known TCM schemes can indeed be obtained.

The classification of labelings presented this paper does not make any assumption on the channel nor on the receiver. Because of this, the presented design methodology can be used to design optimal TCM encoders for other channels as well as for suboptimal (BICM) decoders.

The algorithm introduced in this paper to find all the labelings that need to be tested in an exhaustive search becomes impractical for constellations with more than 16 points. In this case, a suboptimal solution based on an algorithm (inspired by the linearity increasing swap algorithm of \cite[Sec.~IX]{Knagenhjelm96}) that generates a subset of (good) labelings could be devised. This approach could also be combined with the concept of labelings with optimal EP \cite{Wesel01}. The design of such an algorithm is left for further investigation.

\section*{Acknowledgement}\label{Sec:Ack}

The authors would like to thank R. F. H. Fischer for pointing out the equivalence between TCM encoders with encoders optimized for the BRGC and the NBC, and showing how the encoders in \cite{Alvarado10d} and \cite{Ungerboeck82} are related. These observations inspired this full paper. The authors would also like to thank R. D. Wesel for fruitful discussions.


\bibliography{IEEEabrv,references_all}

\begin{thebibliography}{10}
\providecommand{\url}[1]{#1}
\csname url@samestyle\endcsname
\providecommand{\newblock}{\relax}
\providecommand{\bibinfo}[2]{#2}
\providecommand{\BIBentrySTDinterwordspacing}{\spaceskip=0pt\relax}
\providecommand{\BIBentryALTinterwordstretchfactor}{4}
\providecommand{\BIBentryALTinterwordspacing}{\spaceskip=\fontdimen2\font plus
\BIBentryALTinterwordstretchfactor\fontdimen3\font minus
  \fontdimen4\font\relax}
\providecommand{\BIBforeignlanguage}[2]{{%
\expandafter\ifx\csname l@#1\endcsname\relax
\typeout{** WARNING: IEEEtran.bst: No hyphenation pattern has been}%
\typeout{** loaded for the language `#1'. Using the pattern for}%
\typeout{** the default language instead.}%
\else
\language=\csname l@#1\endcsname
\fi
#2}}
\providecommand{\BIBdecl}{\relax}
\BIBdecl

\bibitem{Ungerboeck76}
G.~Ungerboeck and I.~Csajka, ``On improving data-link performance by increasing
  channel alphabet and introducing sequence decoding,'' in \emph{International
  Symposium on Information Theory (ISIT)}, Ronneby, Sweden, June 1976, (Book of
  abstracts).

\bibitem{Ungerboeck82}
G.~Ungerboeck, ``Channel coding with multilevel/phase signals,'' \emph{{IEEE}
  Trans. Inf. Theory}, vol.~28, no.~1, pp. 55--67, Jan. 1982.

\bibitem{Ungerboeck87a}
------, ``Trellis-coded modulation with redundant signal sets {Part I}:
  Introduction,'' \emph{{IEEE} Commun. Mag.}, vol.~25, no.~2, pp. 5--11, Feb.
  1987.

\bibitem{Ungerboeck87b}
------, ``Trellis-coded modulation with redundant signal sets {Part II}: State
  of the art,'' \emph{{IEEE} Commun. Mag.}, vol.~25, no.~2, pp. 12--21, Feb.
  1987.

\bibitem{Biglieri91_Book}
E.~Biglieri, D.~Divsalar, P.~J. McLane, and M.~K. Simon, \emph{Introduction to
  Trellis-Coded Modulation with Applications}.\hskip 1em plus 0.5em minus
  0.4em\relax Macmillan, 1991.

\bibitem{Proakis08_Book}
J.~G. Proakis and M.~Salehi, \emph{Digital Communications}, 5th~ed.\hskip 1em
  plus 0.5em minus 0.4em\relax McGraw-Hill, 2008.

\bibitem{Lin04_Book}
S.~Lin and D.~J. {Costello, Jr.}, \emph{Error Control Coding}, 2nd~ed.\hskip
  1em plus 0.5em minus 0.4em\relax Englewood Cliffs, NJ: Prentice Hall, 2004.

\bibitem{Zehavi92}
E.~Zehavi, ``8-{PSK} trellis codes for a {Rayleigh} channel,'' \emph{{IEEE}
  Trans. Commun.}, vol.~40, no.~3, pp. 873--884, May 1992.

\bibitem{Caire98}
G.~Caire, G.~Taricco, and E.~Biglieri, ``Bit-interleaved coded modulation,''
  \emph{{IEEE} Trans. Inf. Theory}, vol.~44, no.~3, pp. 927--946, May 1998.

\bibitem{Fabregas08_Book}
A.~{Guill\'en i F\`abregas}, A.~Martinez, and G.~Caire, ``Bit-interleaved coded
  modulation,'' \emph{Foundations and Trends in Communications and Information
  Theory}, vol.~5, no. 1--2, pp. 1--153, 2008.

\bibitem{Gray53}
F.~Gray, ``Pulse code communications,'' U.~S. Patent 2\,632\,058, Mar. 1953.

\bibitem{Agrell04}
E.~Agrell, J.~Lassing, E.~G. Str\"{o}m, and T.~Ottosson, ``On the optimality of
  the binary reflected {G}ray code,'' \emph{{IEEE} Trans. Inf. Theory},
  vol.~50, no.~12, pp. 3170--3182, Dec. 2004.

\bibitem{Alvarado11b}
A.~Alvarado, F.~Br\"{a}nnstr\"{o}m, and E.~Agrell, ``High {SNR} bounds for the
  {BICM} capacity,'' in \emph{IEEE Information Theory Workshop (ITW)}, Paraty,
  Brazil, Oct. 2011.

\bibitem{Chang97}
J.-J. Chang, D.-J. Hwang, and M.-C. Lin, ``Some extended results on the search
  for good convolutional codes,'' \emph{{IEEE} Trans. Inf. Theory}, vol.~43,
  no.~6, pp. 1682--1697, Sep. 1997.

\bibitem{Bocharova97}
I.~E. Bocharova and B.~D. Kudryashov, ``Rational rate punctured convolutional
  codes for soft-decision {Viterbi} decoding,'' \emph{{IEEE} Trans. Inf.
  Theory}, vol.~43, no.~4, pp. 1305--1313, July 1997.

\bibitem{Frenger99}
P.~Frenger, P.~Orten, and T.~Ottosson, ``Convolutional codes with optimum
  distance spectrum,'' \emph{{IEEE} Trans. Commun.}, vol.~3, no.~11, pp.
  317--319, Nov. 1999.

\bibitem{Stierstorfer10}
C.~Stierstorfer, R.~F.~H. Fischer, and J.~B. Huber, ``Optimizing {BICM} with
  convolutional codes for transmission over the {AWGN} channel,'' in
  \emph{International Zurich Seminar on Communications}, Zurich, Switzerland,
  Mar. 2010.

\bibitem{Alvarado10d}
A.~Alvarado, L.~Szczecinski, and E.~Agrell, ``On {BICM} receivers for {TCM}
  transmission,'' \emph{{IEEE} Trans. Commun.}, vol.~59, no.~10, pp.
  2692--2702, Oct. 2011.

\bibitem{Fischer11PrivCommun}
R.~F.~H. Fischer, \emph{private communication}, Jan. 2011.

\bibitem{Zhang96_Thesis}
W.~Zhang, ``Finite state systems in mobile communications,'' Ph.D.
  dissertation, University of South Australia, Adelaide, Australia, Feb. 1996.

\bibitem{Zhang94}
W.~Zhang, C.~Schlegel, and P.~Alexander, ``The bit error rate reduction for
  systematic {8PSK} trellis codes by a {G}ray scrambler,'' in \emph{IEEE
  International Conference on Universal Wireless Access}, Melbourne, Australia,
  Apr. 1994.

\bibitem{Gray99_Thesis}
P.~K. Gray, ``Serially concatenated trellis coded modulation,'' Ph.D.
  dissertation, University of South Australia, Adelaide, Australia, Mar. 1999.

\bibitem{Schlegel04_Book}
C.~B. Schlegel and L.~C. Perez, \emph{Trellis and Turbo Coding}, 1st~ed.\hskip
  1em plus 0.5em minus 0.4em\relax John Wiley \& Sons, 2004.

\bibitem{Barry04_Book}
J.~B. Barry, E.~A. Lee, and D.~G. Messerschmitt, \emph{Digital Communication},
  3rd~ed.\hskip 1em plus 0.5em minus 0.4em\relax Springer, 2004.

\bibitem{Du89}
J.~Du and M.~Kasahara, ``Improvements of the information-bit error rate of
  trellis code modulation systems,'' \emph{The Transactions of the IEICE}, vol.
  E~72, no.~5, pp. 609--614, May 1989.

\bibitem{Clark81_Book}
G.~C. {Clark, Jr.} and J.~B. Cain, \emph{Error-correction coding for digital
  communications}, 2nd~ed.\hskip 1em plus 0.5em minus 0.4em\relax Plenum Press,
  1981.

\bibitem{Viterbi89}
A.~J. Viterbi, J.~K. Wolf, E.~Zehavi, and R.~Padovani, ``A pragmatic approach
  to trellis-coded modulation,'' \emph{{IEEE} Commun. Mag.}, vol.~27, no.~7,
  pp. 11--19, July 1989.

\bibitem{Wesel01}
R.~D. Wesel, X.~Liu, J.~M. Cioffi, and C.~Komninakis, ``Constellation labeling
  for linear encoders,'' \emph{{IEEE} Trans. Inf. Theory}, vol.~47, no.~6, pp.
  2417--2431, Sep. 2001.

\bibitem{Wesel12PrivCommun}
R.~D. Wesel, \emph{private communication}, July 2012.

\bibitem{Knagenhjelm96}
P.~Knagenhjelm and E.~Agrell, ``The {H}adamard transform---a tool for index
  assignment,'' \emph{{IEEE} Trans. Inf. Theory}, vol.~42, no.~4, pp.
  1139--1151, July 1996.

\bibitem{Benedetto99_Book}
S.~Benedetto and E.~Biglieri, \emph{Principles of Digital Transmission with
  Wireless Applications}.\hskip 1em plus 0.5em minus 0.4em\relax Kluwer
  Academic, 1999.

\bibitem{Li02}
X.~Li, A.~Chindapol, and J.~A. Ritcey, ``Bit-interlaved coded modulation with
  iterative decoding and {8PSK} signaling,'' \emph{{IEEE} Trans. Commun.},
  vol.~50, no.~6, pp. 1250--1257, Aug. 2002.

\bibitem{Tran06a}
N.~H. Tran and H.~H. Nguyen, ``Signal mappings of 8-ary constellations for bit
  interleaved coded modulation with iterative decoding,'' \emph{{IEEE} Trans.
  Broadcast.}, vol.~52, no.~1, pp. 92--99, Mar. 2006.

\bibitem{Benedetto88}
S.~Benedetto, M.~A. Marsan, G.~Albertengo, and E.~Giachin, ``Combined coding
  and modulation: Theory and applications,'' \emph{{IEEE} Trans. Inf. Theory},
  vol.~34, no.~2, pp. 223--236, Mar. 1988.

\bibitem{Birkhoff77_Book}
G.~Birkhoff and {S. Mac Lane}, \emph{A Survey of Modern Algebra}, 4th~ed.\hskip
  1em plus 0.5em minus 0.4em\relax New York: Macmillan, 1977.

\bibitem{Duvall71}
P.~F. {Duvall, Jr.} and P.~W. {Harley, III}, ``A note on counting matrices,''
  \emph{SIAM Journal on Applied Mathematics}, vol.~20, no.~3, pp. 374--377, May
  1971.

\bibitem{Calderbank87}
A.~R. Calderbank and N.~J.~A. Sloane, ``New trellis codes based on lattices and
  cosets,'' \emph{{IEEE} Trans. Inf. Theory}, vol. IT-33, no.~2, pp. 177--195,
  Mar. 1987.

\bibitem{Zehavi87}
E.~Zehavi and J.~K. Wolf, ``On the performance evaluation of trellis codes,''
  \emph{{IEEE} Trans. Inf. Theory}, vol. IT-33, no.~2, pp. 196--202, Mar. 1987.

\bibitem{Caire98b}
G.~Caire and E.~Viterbo, ``Upper bound on the frame error probability of
  terminated trellis codes,'' \emph{{IEEE} Commun. Lett.}, vol.~2, no.~1, pp.
  2--4, Jan. 1998.

\bibitem{Sone01}
N.~Sone, M.~Mohri, M.~Morii, and H.~Sasano, ``On good convolutional codes with
  optimal free distance for rates 1/2, 1/3 and 1/4,'' \emph{IEICE Trans.
  Commun.}, vol. E84-B, no.~1, pp. 116--119, Jan. 2001.

\bibitem{Hug12_Thesis}
F.~Hug, ``Codes on graphs and more,'' Ph.D. dissertation, Lund University,
  Lund, Sweden, May 2012.

\end{thebibliography}
\bibliographystyle{IEEEtran}

\end{document}